\documentclass{aa}  

\usepackage{graphicx}
\usepackage{txfonts}
\usepackage{natbib}
\usepackage[table,xcdraw]{xcolor}
\usepackage{hyperref}
\hypersetup{colorlinks=true, citecolor=blue}

\usepackage{mhchem}
\usepackage{txfonts}

\newcommand{\mhalo}{\rm M_{\star,\rm halo}}
\newcommand{\mgal}{\rm M_{\star,\rm gal}}
\newcommand{\msun}{\rm M_{\odot}}

\newcommand{\ropt}{\rm r_{\rm opt}}
\newcommand{\rvir}{\rm r_{\rm vir}}
\newcommand{\cielo}{{\small CIELO}}
\newcommand{\tst}{\rm t_{\rm stable}}
\newcommand{\tamr}{\rm t_{\rm AMR}}
\newcommand{\tninety}{\rm t_{90}}

\newcommand{\tdisr}{\rm t^{\rm SHMC1}_{\rm merger}}

\defcitealias{Gonzalez-Jara_2025}{GJ25}

\begin{document}

   \title{The heartbeat of stellar halos: Insights from the stellar halo mass-metallicity relation}
    \author{Jenny Gonzalez-Jara
          \inst{1}\inst{2}\fnmsep\thanks{E-mail: jagonzalez11@uc.cl}
          \and
        Patricia B. Tissera\inst{1}\inst{2}
        \and
        Antonela Monachesi\inst{3} 
          \and 
          Brian Tapia-Contreras\inst{1}\inst{2}
          \and
          Susana Pedrosa\inst{4} 
          \and
          Catalina Casanueva-Villarreal\inst{1}\inst{2}
          \and
          Rosa Domínguez-Tenreiro\inst{5}
          \and 
          Lucas Bignone\inst{4}}
        \institute{Instituto de Astrofísica, Pontificia Universidad Católica de Chile. Av. Vicuña Mackenna 4860, Santiago, Chile.
         \and
    Centro de AstroIngeniería, Pontificia Universidad Católica de Chile. Av. Vicuña Mackenna 4860, Santiago, Chile.
        \and 
    Departamento de Astronomía, Universidad de La Serena, Av. Raúl Bitrán 1305, La Serena, Chile.
    \and 
    Instituto de Astronomía y Física del Espacio, CONICET-UBA, Casilla de Correos 67, Suc. 28, 1428 Buenos Aires, Argentina.
    \and
    Departamento de Física Teórica, Universidad Autónoma de Madrid, E-28049 Cantoblanco, Madrid, Spain.
    }
   \date{Received 8 December 2025 / Accepted 24 May 2026}

 
  \abstract
   {The stellar halo mass–metallicity relation (MZhR) has been observed in nearby Milky Way–mass galaxies and explored in numerical simulations. Recent advances in understanding stellar halo assembly, both observationally and theoretically, motivate the investigation of whether the MZhR is already in place at high redshift.}
   {This work aims to investigate the presence and evolution of the MZhR from redshift $z\approx3.5$ to $z=0$, and to identify when galaxies settle on the present-day MZhR.}
   {We used central galaxies with stellar mass of $\rm log_{10} (\mgal/\msun) \in [9-11]$ from CIELO cosmological hydrodynamical zoom-in simulations. We identified stellar halos, from $z\approx3.5$ to $z=0$, using a dynamical decomposition method, AM–E, focusing on the region between the 1.5 optical radius and the virial radius. The galaxy sample includes stellar halos with $\rm log_{10}(\mhalo/\msun) \in [8-10]$ at $z=0$. We presented halo cardiograms, a novel approach to studying the assembly history of stellar halos. Using them, we defined a stability time ($\tst$) as the first time that the median halo metallicity does not change more than $\pm0.1$~dex with respect to its value at $z=0$.}
   {CIELO stellar halos reproduce the present-day observed MZhR. At $z\approx3.5$, stellar halos already define an MZhR whose slope is similar to the slope at $z=0$. For a fixed stellar halo mass, the metallicity increases $\sim 0.21$~dex from $z\approx3.5$ to $z=0$, reflecting the progressive chemical enrichment provided by the accretion of satellites with diverse masses and different levels of enrichment. When the first stellar halo main contributor (SHMC1) provides a mass fraction at least 20\% higher than the remaining contributors, the stellar halo metallicity is set once SHMC1 is fully disrupted (merger time: $\tdisr$). This yields a clear correlation between $\tst$ and $\tdisr$, with a scatter of $2.2$~Gyr driven by the relative importance of the second and third main contributing satellites. Finally, we provide two observational tracers for $\tst$: $\tninety$ and a stability time from the age-metallicity relation, $\tamr$. Using the $\tst-\tninety$ relation together with $\tst \approx \tdisr$, we infer that if the bulk of the MW's stellar halo was built up by a single major merger, we predict its merger time occurring at $\sim7.5$~Gyr with an uncertainty of 2.8~Gyr estimated via error propagation. While for M31, we inferred a merger time about $1.7$~Gyr suggesting that its stellar halo just reached the MZhR at $z=0$.}
   {Our results suggest that estimating $\tst$ could serve as a proxy for dating the moment at which the stellar halo reaches the present-day MZhR, as well as for dating the last major merger that builds up the stellar halo of galaxies. Additionally, together with an estimation of the merger time of the main contributing satellite, it can provide insights into the relative importance of the second and third contributing satellites.}
   

   \maketitle

\section{Introduction}

Chemical elements heavier than helium, commonly referred as metals in astronomy, provide unique insight into the processes that drive the formation and evolution of galaxies \citep{Freeman_2002,Matteucci_2012}. They  are synthesized in stars through thermonuclear reactions and subsequently expelled into the interstellar medium (ISM), polluting the gas that fuels the formation of new generations of stars \citep{Nomoto_2013}. The abundance of chemical elements, also known as metallicity, is regulated by the interplay between gas inflows, metal recycling via galactic outflows, star formation, mergers and interactions \citep{Maiolino_2019}. As a result, scaling relations between stellar mass and metallicity are naturally expected to arise and trace the physical processes shaping galaxy evolution \citep{Tinsley_79}.

The gas-phase mass-metallicity relation  (MZ$_{\rm g}$R) is one of the most extensively studied scaling relations for galaxies \citep{Lequeux_1979, Tinsley_79}, which stores information on the level of enrichment of star formation regions. The advent of the Sloan Digital Sky Survey \citep[SDSS,][]{Survey_SDSS_2000} provided stellar mass and metallicity measurements in the Local Universe, enabling the determination of the stellar mass-metallicity relation (MZ$_\star$R) for local galaxies \citep{Gallazzi_2005, Kirby_2013}. While the MZ$_{\rm g}$R has been studied for several decades \citep{Lequeux_1979,Tinsley_79,Tremonti_2004,Lee_2006, Maiolino_2008, Torrey_2019,Sanders_2021,Langeroodi_2023}, the MZ$_\star$R has been less explored in extragalactic sources, in part due to observational challenges. The MZ$_\star$R reflects the cumulative fossil record of the star formation and assembly histories of galaxies \citep{Gallazzi_2005,Panter_2008,Kirby_2013,Ma_2015,De_Rossi_2015,Dave_2017,Trussler_2019,Dominguez_Gomez_2023}. Even fewer studies have investigated the stellar halo mass-metallicity relation \citep[MZhR,][]{Deason_2016, Harmsen_2017, DSouza_2018, Murphy_2022}, which is considered to provide insights into the accretion history of their host galaxy \citep{Robertson_2005, Font_2006, Zolotov_2009,Tissera_2013, DSouza_2018, Monachesi_2019, Smercina_2022}.
The local MZhR has been determined by using observations of some nearby spiral galaxies and the Milky Way (MW) \citep{Harmsen_2017, Smercina_2022}. As expected, this relation is consistent with more massive stellar halos exhibiting higher metallicities. Studying stellar halos in external galaxies remains a challenge due to their low-surface brightness, but in the near future the Vera Rubin Survey and later the ELT, will enable to extend the detection of  stellar halos to more galaxies.

Numerical simulations have played an important role in linking stellar halo properties with their accretion histories. \citep{Zolotov_2009,Tissera_2014,Elias_2018,Monachesi_2019,Pulsoni_2020, Canas_2020, Rey_2021, Proctor_2024, Tau_2025, Gonzalez-Jara_2025, Celiz_2025, Vera-Casanova_2025}. They reported that stellar halos were formed by a significant contribution of accreted satellites with an additional in-situ contribution whose relative importance remains debated and is model dependent \citep[e.g.][]{Tissera_2014, Cooper_2015, Monachesi_2019, Tau_2025, Cooper_2025, Wittig_2025}.

\cite{DSouza_2018} investigated the MZhR using Illustris simulations \citep{Vogelsberger_2013_Illustris, Genel_2014_Illustris}, focusing on galaxies with dark matter halo in the range $\rm log_{10} (\rm M_{DM}/\msun) \in [11, 14]$. Illustris galaxies have a large fraction of in-situ stars at large galactocentric distances, which appear to be incompatible with present GHOSTS observations \citep{Harmsen_2017}.
Hence, \cite{DSouza_2018} studied the accreted MZhR and reported that the dominant progenitor builds most of the accreted mass and drives the accreted MZhR. The scatter in this relation was reported to encode information about the mass of the dominant progenitor and, at a given accreted stellar halo mass, reflects the diversity of accretion histories of the stellar halos. They also found that, at a given accreted stellar mass, high metallicity halos tend to have higher fractions of accreted material contributed by one single massive merger event. Similar findings were later reported by \citet{Monachesi_2019} using AURIGA simulations. 

Regarding the stellar halo mass assembly, numerical simulations suggest that  most of the mass in the MW's stellar halo is expected to be contributed by a few massive accretion events \citep{Monachesi_2019, Yun_Pu_2025} along with in-situ stars that are more frequent in the inner stellar halo \citep{Zolotov_2009, Purcell_2010, Tissera_2013, Font_2020}. A simulated reconstruction of the assembly of stellar halos beyond MW–like galaxies was presented in \citet{Gonzalez-Jara_2025} (hereafter GJ25). These authors focused on the outer halo, excluding the inner region of the halos that coexist with the host galaxy, with stellar masses in the range of log$_{10} (\mhalo/\msun) \in [8,10]$ from the \cielo~simulations \citep{Tissera_2025}. Their results indicate that as the stellar halo mass increases, so does the number of significant progenitors, which in general, contribute more than 60\% of the outer halos.

Satellite galaxies that contribute to the formation of stellar halos follow a MZ$_\star$R \citep{Fattahi_2020, Naidu_2022, Khoperskov_2023c, Grimozzi_2024}. Disrupted dwarf galaxies, in particular, tend to be less metal-rich than surviving satellites at fixed stellar mass. This difference reflects their distinct star formation timescales, with surviving systems typically sustaining star formation for longer periods \citep{Tissera_2012}. If the MZ$_\star$R evolves with lookback time, the stellar halos might inherit a similar trend given they are mainly built from accreted stars from dwarf galaxies. 

Recent observations from the CLAUDS and HSC-SSP surveys have enabled the investigation of stellar halo assembly in the redshift range $0.2 \leq z \leq 1.1$ \citep{williams_2024}. These efforts, along with advances in numerical simulations, reveal that low- and high-mass stellar halos exhibit different rates of mass growth \citep{Amorisco_2017, Elias_2018, Tau_2025, Gonzalez-Jara_2025}, largely driven by differences in their accretion histories. If stellar halos evolve through distinct phases of growth and quiescence, it remains to be understood when the MZhR first emerges, how it evolves over time, and what information this relation encodes about the assembly history of stellar halos and their host galaxies. The advancement of observational data prompts us to explore these questions further using simulations which include galaxies of a variety of stellar masses. 

In this paper, we use the \cielo~simulations to investigate the MZhR across cosmic time. We use the set of galaxies and their stellar halos studied by \citetalias{Gonzalez-Jara_2025} and analyze their assembly histories following their progenitors back in time up to $z \approx 3.5$. 
We compared our simulated halos with available observations at $z \sim 0$ by applying a similar procedure and radial aperture. However, to understand the assembly of the stellar halos and the evolution of the MZhR, we focus on the stellar halos beyond the region where it coexists with its host galaxy.
We extend previous numerical studies beyond the MW-mass halos. We aim to understand when a stellar halo of a galaxy reaches the local MZhR and what features of its assembly history are stored in the MZhR, its dispersion, and evolution. We present halo cardiograms, a novel approach to study the assembly of stellar halos, and we propose a way to build observables cardiograms.

The paper is organized as follows. Section~\ref{sec:methods} briefly describes the \cielo~simulations and the definitions adopted throughout this work. Section~\ref{sec:mzhr_z0} compares the MZhR at $z=0$ using current observations and simulations. Section~\ref{sec:MZ_evolution} investigates the stellar halo mass and metallicity from redshift $z\approx 3.5$ to $z=0$. In Section~\ref{sec:heartbeat_sh}, we introduce cardiograms of stellar halos as an approach to study the halo assembly and we identify when the median halo metallicity is stabilized, i.e. does not change more than $\pm$0.1~dex respect to its value at $z=0$. In Section~\ref{sec:observations}, we present observational measurements as proxies to estimate the stability time. Finally, in Section~\ref{sec:discussion} we discuss our results and caveats, followed by a summary in Section~\ref{sec:summary}.

\section{Methodology}\label{sec:methods}
In this section, we introduce the \cielo~simulations (Sec.~\ref{sec:simulations}) used in this work and describe the adopted definitions  (Sec.~\ref{sec:sample}) for identifying stellar halos, their stellar populations, and contributing satellites.

\subsection{CIELO simulations}\label{sec:simulations}

The CIELO simulations are fully cosmological hydrodynamical zoom-in simulations that assume a $\Lambda$ Cold Dark Matter universe  with $\Omega_0$ = 0.317, $\Omega_{\Lambda}$ = 0.682, $\Omega_{B}$ = 0.049, $h$= 0.671, $\sigma_{8} = 0.834$, $n_{s}=0.962$ \citep{Planck2014}. The CIELO galaxies are embedded within zoom-in regions, which include small groups, walls, and filaments \citep{Tissera_2025}. 

The \cielo~suite includes two levels of resolution: L11 with $\rm m_{\rm dm} = 1.36 \times 10^6 \msun \rm h^{-1}$ and $\rm m_{\rm gas} = 2.0 \times 10^5 \msun \rm h^{-1}$ and L12 with $\rm m_{\rm dm} = 1.36 \times 10^5 \msun \rm h^{-1}$ and $\rm m_{\rm gas} = 2.1 \times 10^4 \msun \rm h^{-1}$. These simulations were run using a version of {\small GADGET-3}, an updated version of {\small GADGET-2} \citep{Springel_2003, Springel_2005}. This version includes subgrid models for the multiphase ISM, metal-dependent radiative cooling, stochastic star formation, and energy and feedback by Type Ia and II supernovae (SNIa and SNII, respectively) as described in \citet{Scannapieco_2005, Scannapieco_2006}. The chemical isotopes included in the simulation are the following twelve: H,\ce{^{4} He}, \ce{^{12} C}, \ce{^{14} N}, \ce{^{16} O}, \ce{^{20} Ne}, \ce{^{24} Mg}, \ce{^{28} Si}, \ce{^{32} S}, \ce{^{40} Ca}, \ce{^{56} Fe}, and \ce{^{62} Zn} \citep{Mosconi_2001}. The multiphase and SN feedback models are able to trigger violent galactic mass-loaded winds without introducing mass-scale parameters \citep{Scannapieco_2008}.

The simulated galaxies are identified using a friends-of-friends algorithm \citep[FoF;][]{Davis_1985} and the SUBFIND algorithm \citep{Springel_2001, Dolag_2009}. The merger trees were built using the AMIGA method \citep{Knollmann_2009}.
 
The \cielo~simulations have been used in previous studies to investigate the effects of the environment on infalling satellite galaxies \citep{Rodriguez_2022}, the impact of baryons on the shape of dark matter halos \citep{Cataldi_2023}, metallicity profiles of galaxies \citep{Tapia_2022,Tapia-contreras_2025}, primordial black holes as a possible component of the dark matter content \citep{Casanueva_2024}, the formation channels of stellar halos \citepalias{Gonzalez-Jara_2025}, the mass-metallicity relation of bulges \citep{Munoz-escobar_2025} and metal-loaded outflows in low-mass galaxies \citep{miranda2025}.

\subsection{Sample and definitions}\label{sec:sample}

In this work, we consider the sample introduced in \citetalias{Gonzalez-Jara_2025}, which consists of 28 stellar halos of central galaxies with stellar masses in the range of 
$\log_{10}(\mgal/\mathrm{M_\odot}) \in [9,11]$ and dark-matter halo masses of $\log_{10}(\rm M_{\rm DM}/\mathrm{M_\odot}) \in [10,12]$. The half-mass radii of our galaxies range from $\sim 1$ to $\sim 5$~kpc, and the virial radii, $\rvir$,  from $\sim$ 80 to 240~kpc\footnote{We define $\rvir$, as the radius enclosing the overdensity of $200$ times the critical density of the Universe, including baryons and dark matter. This was done at all analyzed redshift. In our previous work, \citetalias{Gonzalez-Jara_2025}, we used the $\rvir$ calculated by the Rockstar halo finder, which considered only dark matter (see Table 1 of \citetalias{Gonzalez-Jara_2025} which displays  distances in kpc/h).}. Following the same procedure used by \citetalias{Gonzalez-Jara_2025}, the stellar components of each galaxy were identified by using the AM-E method described by \cite{Tissera_2012}, based on the angular momentum and the binding energy of the stellar particles. Briefly, stellar particles that do not belong to the bulge and disk components are considered to be part of the stellar halo.

Following \citetalias{Gonzalez-Jara_2025}, we adopt the same stellar halo definition and focus on the stellar halo region between $1.5\ropt$\footnote{The $\ropt$ is defined as the radius that enclosed about 80 percent of the stellar mass of a galaxy. For our galaxies, 1.5$\ropt$ ranges from 4.23 to 17.02~kpc.} and $\rvir$, which we hereafter define as the outer stellar halo. The resulting outer stellar halos spans stellar masses of $\rm log_{10}(\mhalo/ \msun) \in [8, 10]$. The inner region of stellar halos, where the stellar halo coexists with its host galaxy, will be discussed in a separate paper. 

In order to study the evolution of outer stellar halos, we traced the 28 halos back in time using the merger trees. Stellar masses and metallicities are estimated up to $z\approx3.5$ using the aforementioned dynamical decomposition. We adopted a lower limit for the outer stellar halo mass determined by the numerical resolution of our simulations, considering only outer stellar halos with more than 500 stellar particles at each analyzed redshift.
This corresponds to a stellar mass limit of $\sim 5 \times 10^6~\msun$ and $\sim 5 \times 10^7$~$\msun$ for the resolution levels L12 and L11, respectively.

We classified present-day stellar halo stars into three categories based on their formation site: in-situ, endo-debris, and ex-situ. A detailed characterization of these populations is given in \citetalias{Gonzalez-Jara_2025}. Briefly, in-situ stars were born bound to the central galaxy. Endo-debris stars were mainly formed during wet mergers, bound to a subhalo within the virial radius of the host galaxy. Ex-situ stars were born in another galaxy, located outside the virial radius of the host galaxy, and were later accreted. Since both ex-situ and endo-debris stars originate from accreted material, we define the accreted stellar component as the combination of these two populations.

We define the stellar halo main contributors (SHMCs) as the satellite galaxies that contribute the highest stellar mass to the accreted outer stellar halo, including both ex-situ and endo-debris stars. We rank them according to their relative contribution (SHMC1, SHMC2, etc.). The infall time, $\rm t_{\rm infall}$ is defined as the lookback time of the last snapshot before the satellite crosses the virial radius of the central galaxy. If a satellite enters more than once, then we consider the first time it does so. The merger time, $\rm t_{\rm merger}$, is defined as the lookback time of the snapshot when the satellite is no longer recognized as a separate structure and is considered bound to the central galaxy, according to the merger tree.

\section{The local stellar halo mass metallicity relation}\label{sec:mzhr_z0}

\begin{figure}
    \centering
    \includegraphics[width=0.45\textwidth]{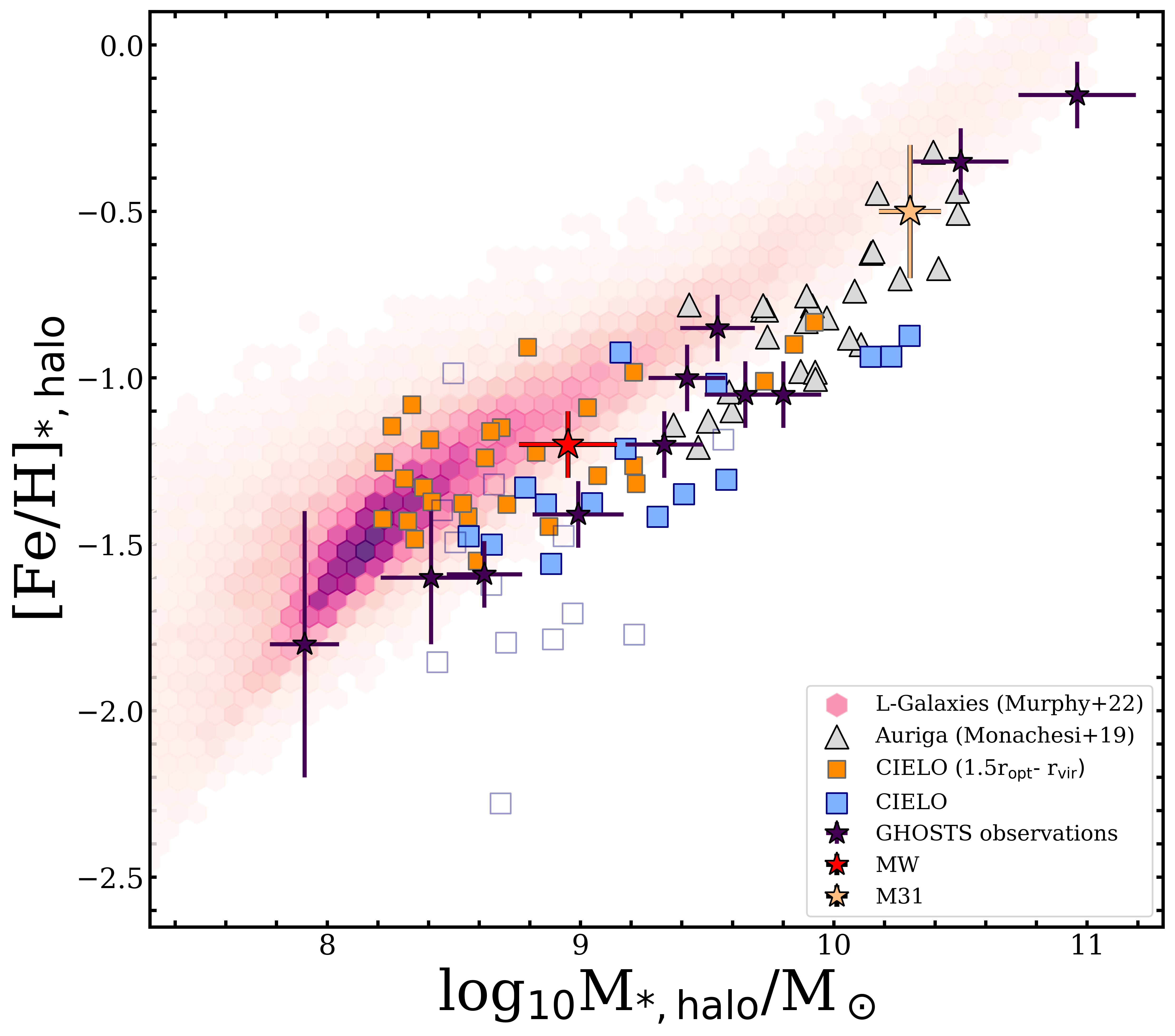}
    \caption{The mass-metallicity relationship for stellar halos at $z=0$. Observational data from GHOSTS HST survey \citep{Harmsen_2017,Jang_2020,Smercina_2022} are represented by stars.
    Estimates for the accreted stellar halo mass and metallicity measured at 30 kpc are displayed for the Auriga galaxies \citep[gray triangles from][]{Monachesi_2019} and the semi-analytic L-Galaxies \citep[pink hexbin from][]{Murphy_2022}. For \cielo~galaxies (blue squares) the metallicity is measured at $6\rm r_{\rm eff}$ (see Sec.~\ref{sec:mzhr_z0} for details) and halos with fewer than 50 particles within the aperture are shown as open squares. In addition, we show the \cielo~stellar halos measured between $1.5\ropt$ and $\rvir$ (orange circles), an aperture that encompasses all stellar populations populations (in-situ, endo-debris and ex-situ stars).
    }
    \label{fig:MhZR_z0}
\end{figure}

As a first step, we calculated the local \cielo~MZhR and compared it with available observations. 
Figure~\ref{fig:MhZR_z0} presents  the observed MZhR at $z=0$ from the GHOSTS survey \cite[black stars,][]{Harmsen_2017,Smercina_2022} and predictions for the accreted stellar halo component from simulations and a semi-analytical model \citep[][pink hexbins]{Murphy_2022}. In addition, we display the median metallicity of the outer stellar halo (orange squares) considering both the accreted and in-situ populations. These estimations correspond to the stellar populations we will focus on this work. Hence, we include them to have a reference of their level of enrichment.

Stellar metallicities reported by \citet{Harmsen_2017} were inferred along the projected minor axis of the disk at a galactocentric radius of 30 kpc, using red giant branch stars in nearby spiral galaxies. The total stellar halo mass was estimated as three times\footnote{The factor of three is derived from comparisons with theoretical models \citep{Bullock_2005, Bell_2017}} the mass enclosed between two ellipses at minor axis radii of 10 and 40 kpc. Figure~\ref{fig:MhZR_z0} includes the six GHOSTS galaxies from \citet{Harmsen_2017} and five additional GHOSTS observations \citep{Jang_2020, Smercina_2022, Gozman_2023} extending the original sample from \citet{Harmsen_2017}. The linear regression fitted to the GHOSTS sample is $\rm [Fe/H]_{30kpc}=0.58\times log_{10}M_{10-40kpc}-6.5$, the slope is shallower compared to the reported in \citet{Harmsen_2017} ($\rm m=0.7$), which was estimated using orthogonal distance regression applied to eight galaxies (six GHOSTS galaxies in addition to MW and M31). To mimic these observational constraints, \citet{Monachesi_2019} using Auriga simulations, measured the median metallicities at 30 kpc along the minor axis computed in 15$^\circ$ projected wedges on the z-coordinate (perpendicular to the disk plane), with diametrically opposed wedges combined to increase numerical resolution \citep[see][]{Monachesi_2016b}. The halo stellar mass is computed as three times the stellar mass beyond 1.5 optical radii along the major axis and 10 kpc above the disk plane. They estimated both quantities by selecting only the accreted component, i.e., all stars formed in satellites, regardless of whether the satellite was inside or outside the virial radius (gray triangles). Figure~\ref{fig:MhZR_z0} also shows stellar halos from the semi-analytic model L-Galaxies 2020-MM \citep[pink hexagons,][]{Murphy_2022}, which are assembled only from accreted satellites, thereby containing exclusively ex-situ stars by definition. This model includes thousands of MW analog galaxies, and their metallicities are rescaled at 30kpc using a linear negative radial profile, calibrated using the mean slope of the metallicity profile from AURIGA simulations \citep{Monachesi_2019}. For the CIELO stellar halos, we compute the stellar mass following the same procedure as \citet{Monachesi_2019}. However, to estimate the metallicity, instead of adopting a fixed aperture at 30 kpc that was used in previous works for MW-like galaxies, we use an aperture that scales with galaxy size since our stellar mass range is wider and 30 kpc could introduce spurious results. For this purpose, we consider that the Milky Way has an effective radius of $\sim 5$ kpc \citep[the mean of the values reported in Table 4 by ][]{Molla_2019}, which implies that 30 kpc corresponds approximately to $6 \rm r_{\rm eff}$. Hence, we adopt $6 \rm r_{\rm eff}$ along the minor axis as our criterion. Motivated by observations we include in-situ stars in our estimations, as separating accreted and in-situ populations is not straightforward in observational data. The observational values reported typically include all stellar populations at these distances. Nevertheless, we note that the contribution of in-situ stars at these galactic distances is, in median value, about 15\%, which increases the metallicity only by $~0.02$~dex up to $0.04$~dex. Likewise, other works have shown that the in-situ contribution at these distances is rather negligible \cite{Monachesi_2019}.

Fig.~\ref{fig:MhZR_z0} shows that the MZhR is reproduced by CIELO stellar halos spanning a broad range of galaxy stellar masses (Sec.~\ref{sec:sample}). GHOSTS survey, Auriga simulations, and the semi-analytic model L-Galaxies correspond to spiral, MW mass-size ones. These galaxies might exhibit different formation histories leading, for example, to differences in the strength of spiral arms and the presence or absence of stellar bars (see e.g., \citet{Fragkoudi_2025}. Our CIELO sample includes a broader morphological diversity, including dwarf elliptical, spiral, and some barred galaxies. Moreover, we did not restrict our selection to MW mass-size galaxies (see Section~\ref{sec:sample}). L-Galaxies exhibit a higher normalization than both the GHOSTS observations and the CIELO simulations. Several factors may contribute to this difference, including the adopted chemical enrichment prescriptions (e.g., the delay time distribution and the nucleosynthetic yields) and the rescaling used to estimate metallicities at 30 kpc, which assumes a negative radial metallicity profile. This rescaling is necessary because the semi-analytic formalism models a homogeneous spatial component, rather than particles, and therefore metallicities at fixed galactocentric distances cannot be measured directly.

Different apertures to measure the metallicity of stellar halos with the corresponding selection of different stellar populations yield variations in the median metallicity. In the case of the \cielo~stellar halos, they have negative metallicity gradients, with the outer region being less metal-poor than the inner one in agreement with previous results \citep{Bullock_2005,Tissera_2014}. The wide range of galaxy sizes motivated us to adopt an aperture that scales with galaxy size, rather than using a fixed aperture at 30 kpc. Moreover, as mentioned above, we study the stellar halo region between $1.5\ropt$ to $\rvir$ (orange squared in Fig.~\ref{fig:MhZR_z0}) including all stellar halo populations (in-situ, ex-situ and endo-debris stars) contained in it. This definition extends to inner regions that contain more metal-rich stars and might include in-situ stars, increasing the metallicity with respect to the $6\rm r_{\rm eff}$ aperture (blue squares) about $0.12^{0.17}_{0.04}$~dex(median value, 25th/75th percentiles as subscript/superscript, respectively), as shown in Fig.~\ref{fig:MhZR_z0}. The $1.5\ropt$ to $\rvir$ aperture adapts to the different masses and sizes of our sample from $z=0$ to $z\approx3.5$ and allows us to study the physical processes involved in the assembly of stellar halos. In the future, if other observational apertures are available, we could explore them to make detailed comparisons. We highlight that even using the two different apertures, the 30~kpc or the region between $1.5\ropt$ and $\rvir$, stellar halos follow a mass-metallicity relation. 

In \citetalias{Gonzalez-Jara_2025}, we showed that outer stellar halos in \cielo~galaxies at $z=0$ are populated mainly by accreted stars, i.e., ex-situ and endo-debris stars ($\sim$80\%, in a median value), which sets the metallicity of the halo at $z=0$. In-situ stars shift the median halo metallicity within a standard deviation of $\sigma = \pm 0.03~\rm dex$. With the exception of three galaxies (P7-7805, LG1-2208, and P4-0000) that have in-situ mass fractions ranging from 25 to 40\% at $z=0$, in which cases the in-situ stars increases the median accreted metallicity $\sim$~0.03 dex up to 0.14~dex. In this work, we study the outer halo including all stellar halo populations (in-situ, ex-situ, and endo-debris stars) contained in it. 

In summary, \cielo~stellar halos follow the observed MZhR at $z=0$ (Fig.~\ref{fig:MhZR_z0}) considering either the aperture comparable to GHOSTS observations (blue squares, $\displaystyle \rm [Fe/H]_{6\rm r_{\rm eff}} = 0.32\times \rm log_{10}\mhalo -4.21$) or the outer halo (orange squares, $\displaystyle \rm [Fe/H]_{1.5\ropt-\rvir} = 0.24\times \rm log_{10}\mhalo -3.31$). We estimated bootstrap errors for both slopes of 0.1 dex using a 5000 resampling. The simulated MZhR for the outer halos,  which includes in-situ, endo-debris and ex-situ stars, has a standard deviation of 0.19 dex. In the following sections, we explore whether the MZhR of the outer halos evolves with time.
 
\section{Stellar halo mass metallicity from $z=3.5$ to $z=0$}\label{sec:MZ_evolution}

In this section, we investigate the stellar halo mass and metallicity over time. We address the following questions: Is the MZhR already in place at high redshift? If so, does it evolve with time? And how does this evolution proceed? We emphasize that the mass and metallicity are estimated at each redshift following the procedure described in Sec.~\ref{sec:simulations}, within the radial range from 1.5$\ropt$ to $\rvir$. This region includes all the stellar halo populations (i.e., in-situ, endo-debris, and ex-situ stars). 

\begin{figure*}
    \centering
    \includegraphics[width=1\linewidth]{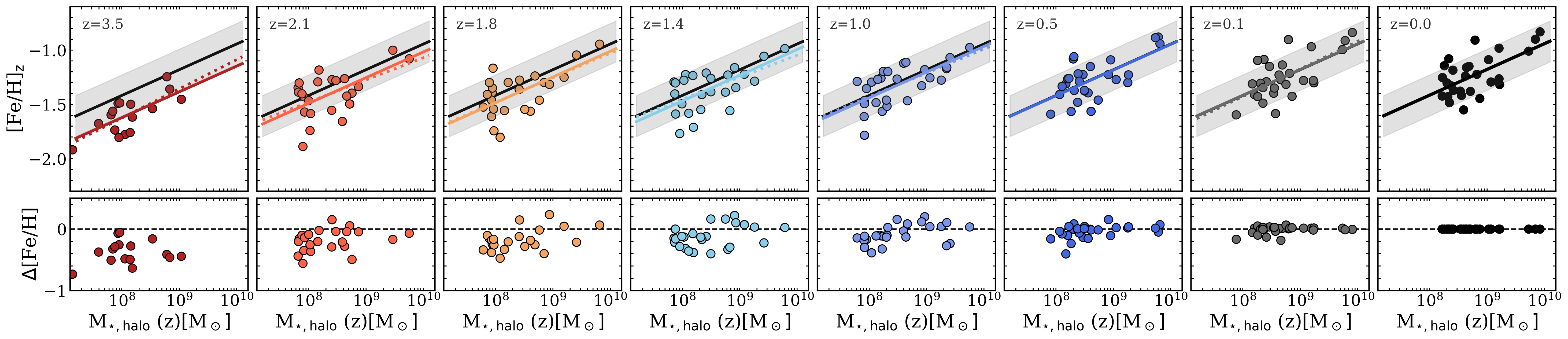}
    \caption{Top panels: Stellar halo mass and metallicity at different redshifts (indicated in each panel). The black line and gray region show the MZhR at $z=0$ and its 1$\sigma$ dispersion. The solid colored lines represent a fit with fixed slope from $z=0$, minimizing the y-intercept at that redshift. The dashed colored line is a linear fit at each time. Bottom panels:    
    Difference in metallicity with respect to $z=0$, $\Delta \rm [Fe/H]=[Fe/H]_z-[Fe/H]_{z=0}$, as a function of the stellar halo mass at the same redshift, $\rm M_{\star,{halo}}(z)$. Negative $\Delta$ values indicate lower metallicities relative to the final values at $z=0$. 
    Only stellar halos with more than 500 stellar particles are displayed according our numerical resolution criteria.}
    \label{fig:summary_MZhR}
\end{figure*}

Figure~\ref{fig:summary_MZhR} displays a well-defined relation between the stellar halo mass and metallicity from $z=3.5$ to $z = 0.0$ for the analyzed simulated halos and the stellar halo of the same galaxy identified along the main branch of the merger tree, at each redshift. The rightmost panel shows the full sample of 28 stellar halos at $z=0$ (also shown  in Fig.~\ref{fig:MhZR_z0} (orange squares). As we move to higher redshift, because of our numerical resolution criterion, some halos are not longer included in the analyzed sample. For example, the leftmost panel includes only 16 halos at $z=3.5$ (see Sec.~\ref{sec:sample}). Each panel shows the \cielo~MZhR at $z=0$ (black line), with a shaded gray region indicating its standard deviation ($\sigma = $0.19~dex). The bottom panel presents the change in metallicity at a given $z$ relative to the present-day value, defined as $\Delta \rm [Fe/H]=[Fe/H]_z-[Fe/H]_{z=0}$. Positive $\Delta$ values can arise when galaxies undergo major mergers because the virial and optical radii may be affected and depending on the relaxation timescale of the event, they required some time to reach again dynamical equilibrium, introducing perturbations in the estimations of stellar halo mass and metallicity due to the redistribution of stars from the inner to the outer halo with respect to the $z=0$ values. It might also be caused by the system moving from the outer to the inner halo. In summary, the dispersion at each time is set by the evolutionary track of each stellar halos.

The lower panel of Fig.~\ref{fig:summary_MZhR} shows that $\Delta \rm[Fe/H]$ decreases towards $z=0$, indicating that stellar halos become progressively enriched, as expected. Low-mass stellar halos primarily accrete material from low-mass galaxies with low star-formation rates (see for example figure 8 and figure 12 from \citetalias{Gonzalez-Jara_2025}) and weak gravitational potential wells, which are less prompt to a retain metals \citep{miranda2025}. As a consequence, they contribute with low metallicity stellar population compared to the more massive satellites accreted by  high-mass stellar halos,  in agreement with previous works  \citep[e.g.][]{Gomez_2010, DSouza_2018, Fattahi_2020}.  Hence, we find that the assembly of stellar halos establishes a well-defined MZhR by at least $z\approx3.5$. We cannot go beyond this redshift because we do not have enough well-resolved stellar halos to make a reliable estimation of the MZhR.

Our simulation predicts only a weak evolution of the MZhR from $z=3.5$ to $z=0.0$, indicating that stellar halos become progressively more enriched over time. To quantify this evolution, we explored two approaches: by performing, at each redshift, a linear fit with both slope and intercept treated as free parameters (dotted line in Fig.~\ref{fig:summary_MZhR}), and  by fixing the slope to that of the $z=0$ MZhR and determining, at each redshift, the intercept that minimizes the residuals (solid line in Fig.~\ref{fig:summary_MZhR}). Both approaches yield consistent results, with differences in the RMS below 1\% at any analyzed redshift. Adopting the fixed-slope method, the MZhR evolves in parallel to MZhR at $z = 0$. Hence, the halo metallicity increases by $\approx 0.21$~dex between $z\approx3.5$ and 0.0, reflecting a gradual enrichment of stellar halos.

\begin{figure}
    \includegraphics[width=1\linewidth]{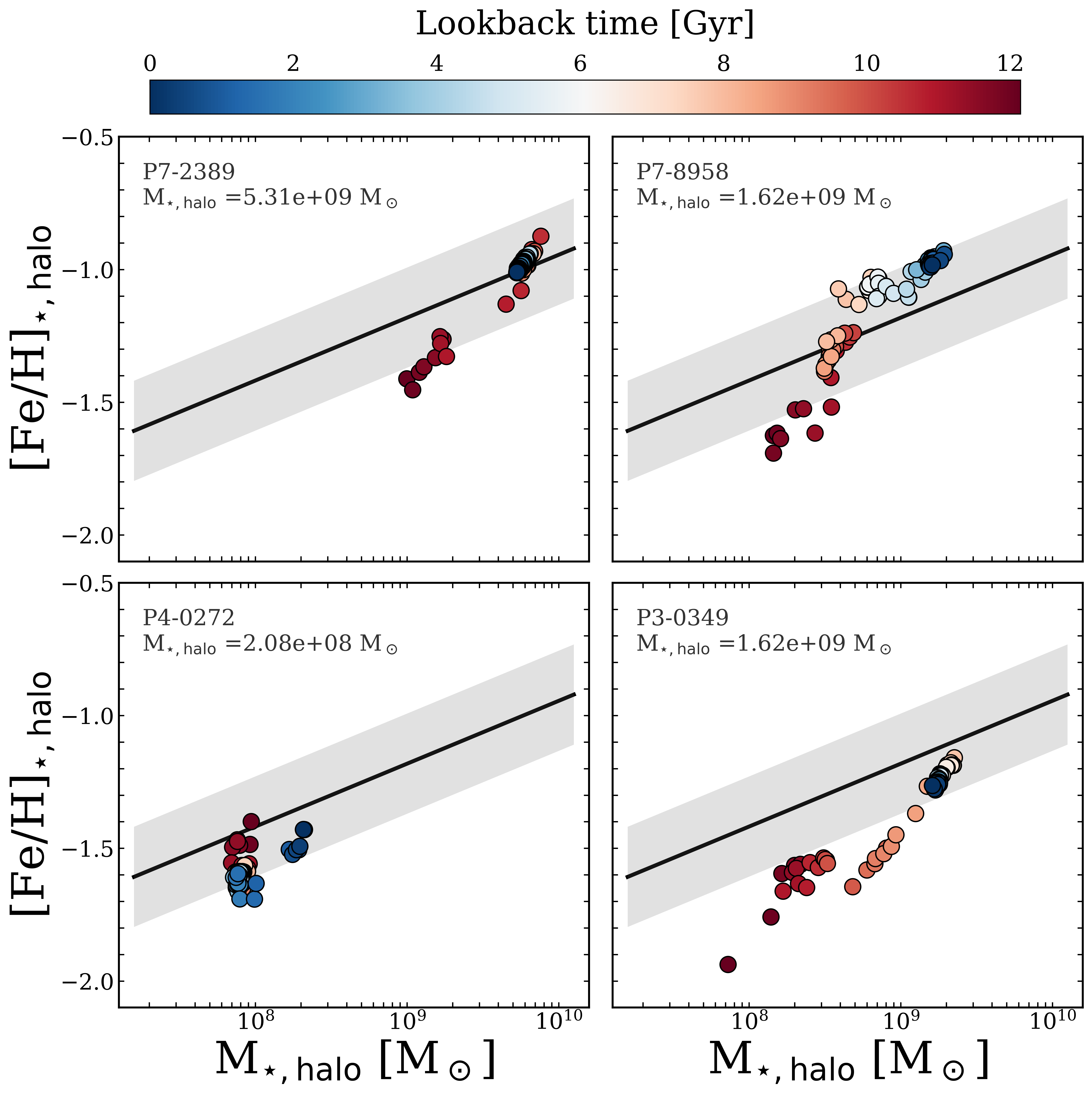}
    \caption{Stellar halo mass and metallicity of four representative galaxies, color-coded by lookback time. The \cielo~MZhR at $z=0$ and 1$\sigma$ dispersion are shown as a black and shaded gray region, respectively. \cielo~galaxy ID and stellar halo mass are displayed in the upper-right corner of each panel.}
    \label{fig:example_paths}
\end{figure}

For illustration purposes, Figure~\ref{fig:example_paths} displays the evolutionary tracks for four stellar halos in the mass-metallicity plane color-coded by lookback time. The left column shows a high-mass stellar halo ($\mhalo = 5.3 \times 10^9~\msun$) at the top and a low-mass halo ($\mhalo = 2.1 \times 10^8~\msun$) at the bottom, while the right column presents two stellar halos with similar masses ($\mhalo = 1.6 \times 10^9~\msun$). The black line and shaded gray region correspond to the MZhR at $z=0$ and its standard deviation, respectively. 
As can be seen from this figure, each stellar halo, regardless of their mass, traces a unique evolutionary path in the mass–metallicity plane. The stellar mass is important to determine the global metallicity of stellar halos (left panels of Fig.~\ref{fig:example_paths}) but other factors can produce differences in the metallicity at a fixed stellar halo mass (right panels of Fig.~\ref{fig:example_paths}). For example, mass and the infall and disruption times of the contributing satellites regulate the chemical enrichment of the halo \citep[see also ][]{Tau_ageFeH_2025}. In the next section, we further investigate the evolutionary tracks of the stellar halos in a more statistical way and in relation to their assembly histories.

\section{The heartbeat of stellar halos}\label{sec:heartbeat_sh}

In this section, we adopt an analogy inspired by the medical field, drawing a parallel between electrocardiograms and the assembly of stellar halos. In cardiology, electrocardiograms record the electrical activity of the heart \citep{Krikler_1987} and reveal irregularities in cardiac rhythm \citep{Fye_1994} or the absence of heartbeats when the signal flattens. Similarly, we present cardiograms that describe the assembly of stellar halos, with the heartbeats represented by variations in their median metallicity and stellar mass. A flat line indicates a quiescent or stable time, where the median metallicity or stellar halo mass do not change significantly within a threshold of $\pm$0.1~dex\footnote{This threshold falls within the standard deviation of the MZhR at $z=0$ ($\sigma =0.19$~dex). Slight variations do not affect our results.} of its value at $z=0$.

\begin{figure*}
    \centering
    \includegraphics[width=1\linewidth]{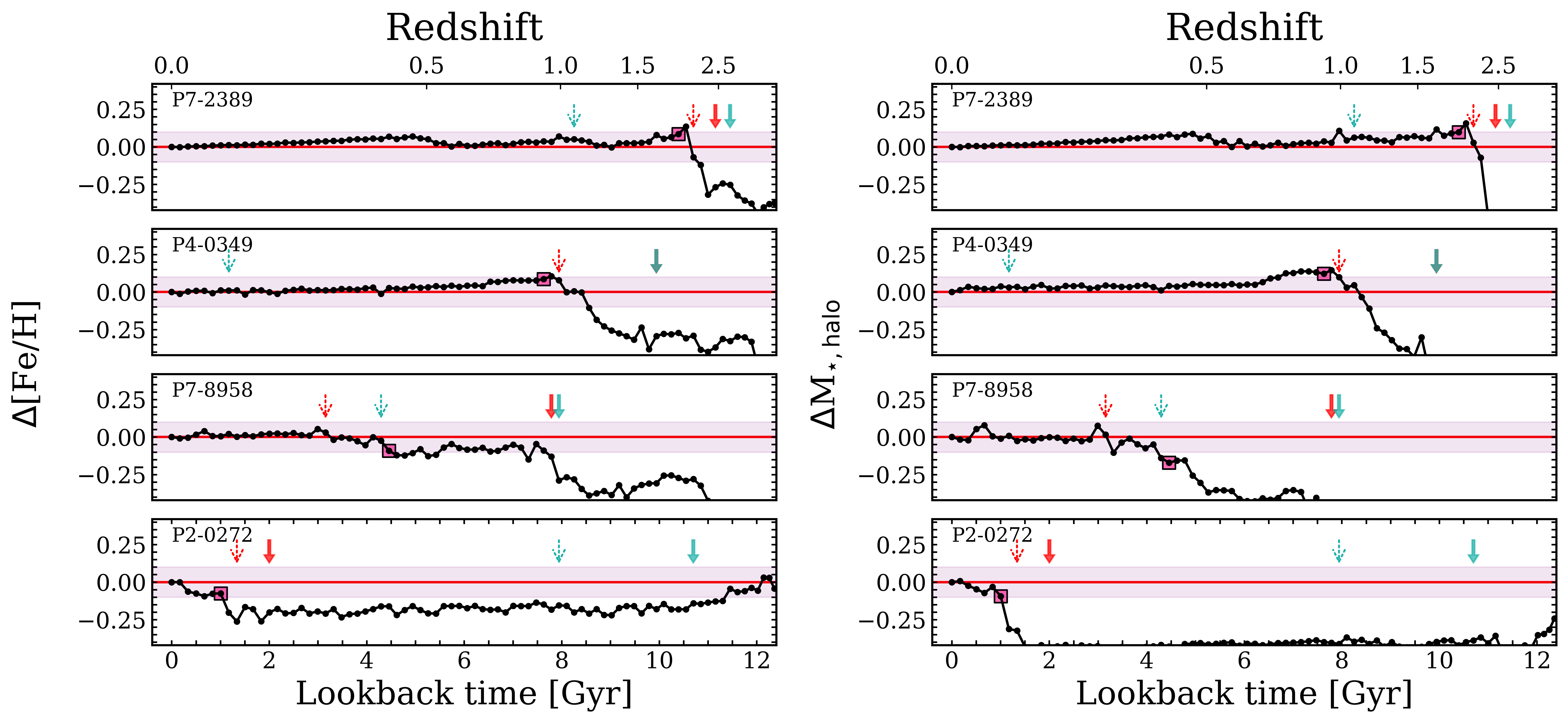}
    \caption{Variations in metallicity (left column) and stellar halo mass (right column) relative to their values at $z=0$ as a function of lookback time. The arrows indicate the main stellar contributors: SHMC1 (red) and SHMC2 (light blue) at infall (solid) and merger
    (dotted) times. The pink shaded region marks differences within $\pm$0.1~dex from $z=0$ of the stellar mass or metallicity. The pink square represent the stability time defined in Sec.\ref{sec:heartbeat_sh}.}
    \label{fig:example_cardiogram}
\end{figure*}

To illustrate this new approach, Figure~\ref{fig:example_cardiogram} presents the cardiograms for the same four stellar halos shown in Fig.~\ref{fig:example_paths}. The cardiograms for the remaining 23 stellar halos\footnote{For the following results, we analyzed 27 stellar halos because one galaxy, LG1-4469, has a surviving SHMC1, therefore it was not considered because we can not estimate a disruption time for their SHMC1.} are displayed in the Appendix~\ref{appendix:all_cardiograms}. The left column shows variations in the median stellar halo metallicity, $\Delta \rm [Fe/H]$ as defined in the previous section. The right column displays changes in the stellar halo mass, defined as $\Delta \rm M_{\star, \rm halo} = log_{10}(M_{\star, \rm halo (z)}) - log_{10}(M_{\star, \rm halo(z=0)})$, as a function of the lookback time. Negative values of $\Delta \rm [Fe/H]$ or $\Delta \rm M_{\star, \rm halo} $ indicate that at the corresponding lookback time, the stellar halo is more metal-poor (left panel) or less massive (right panel) than at its final state at $z=0$. The pink shaded region marks the epochs when the mass or metallicity is within $\pm$0.1 dex of its final value at $z=0$. The arrows denote the infall time (solid) and the merger time (dotted) of SHMC1 (red) and SHMC2 (light blue).

The growth of stellar mass and metallicity follows broadly similar trends with time, reflecting their connection through star formation, stellar evolution, and galaxy assembly. Variations in metallicity and mass are shown since $z\approx3.5$ moving towards $z=0$ (decreasing lookback time). Although the overall trend shows an increase in both stellar mass and stellar metallicity, the cardiograms also display temporary decreases at some times. These variations arise from a combination of factors.
On the one hand, the accretion of SHMCs can introduce metal-poor stars into the outer halo, producing temporary decreases in the median stellar metallicity. This can occur because of two reasons. First, the accreted satellites tend to be less massive and hence bring low-metallicity stars \citep{Munoz-escobar_2025} and secondly, tidal stripping tends to remove the outer, more metal-poor layers of satellite galaxies first, depositing them in the outer halo. On the other hand, the redistribution of particles within the halo can also affect the measured properties.
Because of the orbital configuration, the satellites can have a trajectory, which leads them to the very central regions, and hence, disappears from the outer halo reducing the level of enrichment.
Finally, our definition of the outer stellar halo depends on an inner boundary, the optical radius, which changes at each snapshot as the galaxy evolves and may also be affected by major mergers. These processes produce the small fluctuations observed in the cardiograms.  
In summary, variations in stellar mass and metallicity reflect the dynamical evolution of the stellar halo, driven by the main accretion events and by the adopted outer halo definition. This definition is comparable to observational approaches, where defining the inner boundary remains challenging.

Cardiograms show that galaxies approach a metallicity plateau at a given epoch. We define the stability time ($\tst$) as the first lookback time, traced backward from $z=0$, when the median halo metallicity reaches a value within $\pm$0.1~dex respect to its value at $z=0$ (pink shades). This time is indicated by a pink square in Fig.~\ref{fig:example_cardiogram}. The stability time is defined from the metallicity cardiogram because even when the halo reaches this quiescent phase in global metallicity, it does not imply a complete cessation of satellite accretion, and hence, of the growth in mass. In fact, the stellar halo mass at $\tst$, accounts for $56^{63}_{47}\%$ of its value at $z=0$ (median; lower and upper indices denote the 25th-75th percentiles, respectively). A further $12^{20}_{9}\%$ of stars were in the inner halo or {host galaxy at $\tst$. Some stars can temporally populate the inner halo or galaxy at $\tst$, because they belong to a satellite undergoing disruption at $\tst$, which is no longer identified by the SUBFIND algorithm, resulting in debris deposited in the outer halo at $z=0$. While other stars are dynamically heated by mergers or interactions, ending up in the stellar halo by $z=0$. A small fraction ($0.39^{3.77}_{0.03}\%$) of stars were born after $\tst$ and ended up in the stellar halo at $z=0$. Finally, $25^{36}_{12}\%$ of halo stars at $z=0$ are accreted later than $\tst$.However, their metallicity exhibits a median deviation of only $\sim -0.02$ dex from the $z=0$ value. Therefore, this additional stellar mass does not significantly alter the median metallicity established at $\tst$.

The increase in metallicity in the stellar haloes is primarily driven by the accretion of dwarf galaxies, which follow a MZ$_\star$R (see \citet{Munoz-escobar_2025}, for a discussion of the MZR of disrupted satellites in CIELO simulations). Satellites can experience ongoing star formation after infalling  \citep{Tissera_2014, Slater_2014, Geha_2017}. These stars represent an important formation channel in stellar halos, accounting for $\sim 30\%\rm$ of the accreted halo mass at $z=0$, and are slightly more metal-rich and younger than the ex-situ stars, further enhancing the overall halo metallicity \citepalias{Gonzalez-Jara_2025}. Hence, the progressive contribution of the satellite galaxies leads to a systematic enrichment of the stellar halo over time, as shown in Fig.~\ref{fig:summary_MZhR} and Fig.~\ref{fig:example_paths}. 

Figure~\ref{fig:cardiogram_summary} compiles all halo cardiograms by shifting the time to $\rm t - \tdisr$, where $\tdisr$ is the merger time of SHMC1 that ranges from 0.5 to 11 Gyr depending on the assembly path of each stellar halo. Before SHMC1 is fully disrupted, $\Delta \rm M$ and $\Delta \rm [Fe/H]$ show substantial scatter with median values of $-0.41^{-0.19}_{-0.54}$ and $-0.17^{-0.07}_{-0.29}$ dex, respectively. After  $\tdisr$, the scatter drops significantly to $0.002^{0.04}_{-0.03}$ and $0.01^{0.03}_{-0.004}$ dex, respectively. Therefore, the main contributing satellite plays a key role in setting the stability time, from which the metallicity of stellar halos cease to increase significantly. This agrees with previous works, supporting the idea that the main contributor of the stellar halo is determining MZhR at $z=0$ \citep{Harmsen_2017, DSouza_2018, Monachesi_2019}. The stability time serves as a proxy for the moment when a stellar halo reaches the present-day MZhR, within the standard deviation. 

\begin{figure}
    \centering  \includegraphics[width=0.45\textwidth]{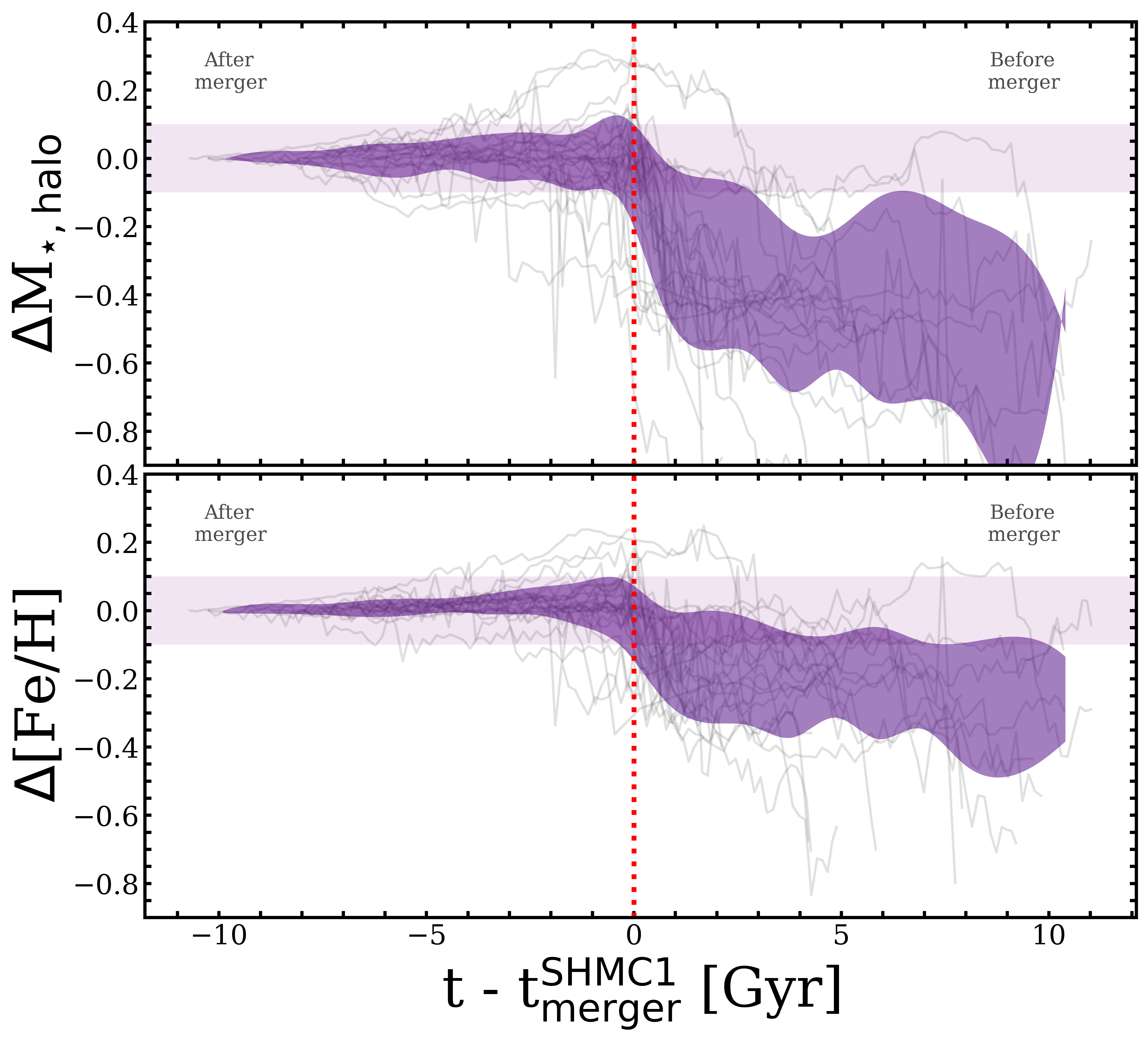}
    \caption{Variations in stellar halo mass (upper panel) and metallicity (lower panel) as a function of lookback time, offset by the merger time of their SHMC1. The purple shaded region represents $\pm1$ standard deviation (16th and 84th percentiles) of the total sample. The pink shaded region indicates a $\pm$0.1 dex variation in mass (upper panel) or metallicity (lower panel) relative to $z=0$.}
    \label{fig:cardiogram_summary}
\end{figure}

Figure~\ref{fig:corr_tst_tdis} shows a strong correlation between $\tst$ and $\tdisr$ (Spearman coefficient $\rm r = 0.72$ and p-value $<2\rm e-05$) with a scatter of $\sigma=2.2$. The color scale quantifies the importance of SHMC1 respect to the second and third main contributor, i.e. $\Delta \rm f_{\rm dom} = f^{SHMC1}_{mass} - f^{SHMC2+SHMC3}_{mass}$. Higher values of $\Delta \rm f_{\rm dom}$ correspond to halos dominated by a single massive accretion event. 

Stellar halos near the one-to-one line tend to assembly most of its stellar mass by the accretion of the SHMC1. They typically reach chemical stability when the disruption of SHMC1 is completed\footnote{We note that if a satellite is assumed to be part of the stellar halo as soon as it gets into the virial radius of the halo, then time of infall should be considered as a reference time. In this work we follow the satellite as a separate structure until it can be longer identified. Hence the contribution of stars to the stellar halo is followed as the satellite is disrupted.}, suggesting that this marks the end of significant enrichment, as shown by the drop in $\Delta \rm [Fe/H]$ in the halo cardiograms (see  Fig.~\ref{fig:cardiogram_summary}). Stellar halos above the one-to-one relation reach $\tst$ before SHMC1 has completely merged, indicating that other contributors  set the level of enrichment of the halo at early time. In these cases, we find that  SHMC2 and SHMC3 together are the ones responsible of this. For halos below the one-to-one line, the contributions of SHMC2 and SHMC3 are also relevant but they contributed at a more recent time in comparison to SHMC1.

In summary, for high  $\Delta \rm f_{\rm dom}$, the stellar halo is mainly assembled through a single dominant accretion event that sets its overall metallicity, contributed by more than 40 percent on average, in agreement with \citet{DSouza_2018}. In such cases, $\tst$ closely matches $\tdisr$, making $\tst$ a useful proxy for constraining the merger time of the last major accretion event that builds up the stellar halo. We recall that merger time is defined as the epoch when the satellites is fully disrupted. In contrast, when $\Delta \rm f_{\rm dom}$ is small or negative, SHMC2 and SHMC3 contribute more substantially to the metallicity of the stellar halo, driving the scatter ($\sigma = 2.2$~Gyr) in the $\tst$–$\tdisr$ plane. Depending on the location on the $\tst$–$\tdisr$ plane the relative time when the contributions occurred could be deduced. Therefore, we suggest that the $\tst$–$\tdisr$ plane stores information on the importance and the chronology of the multiple SHMCs to the formation of the stellar halos.

\begin{figure}
    \centering  \includegraphics[width=0.9\linewidth]{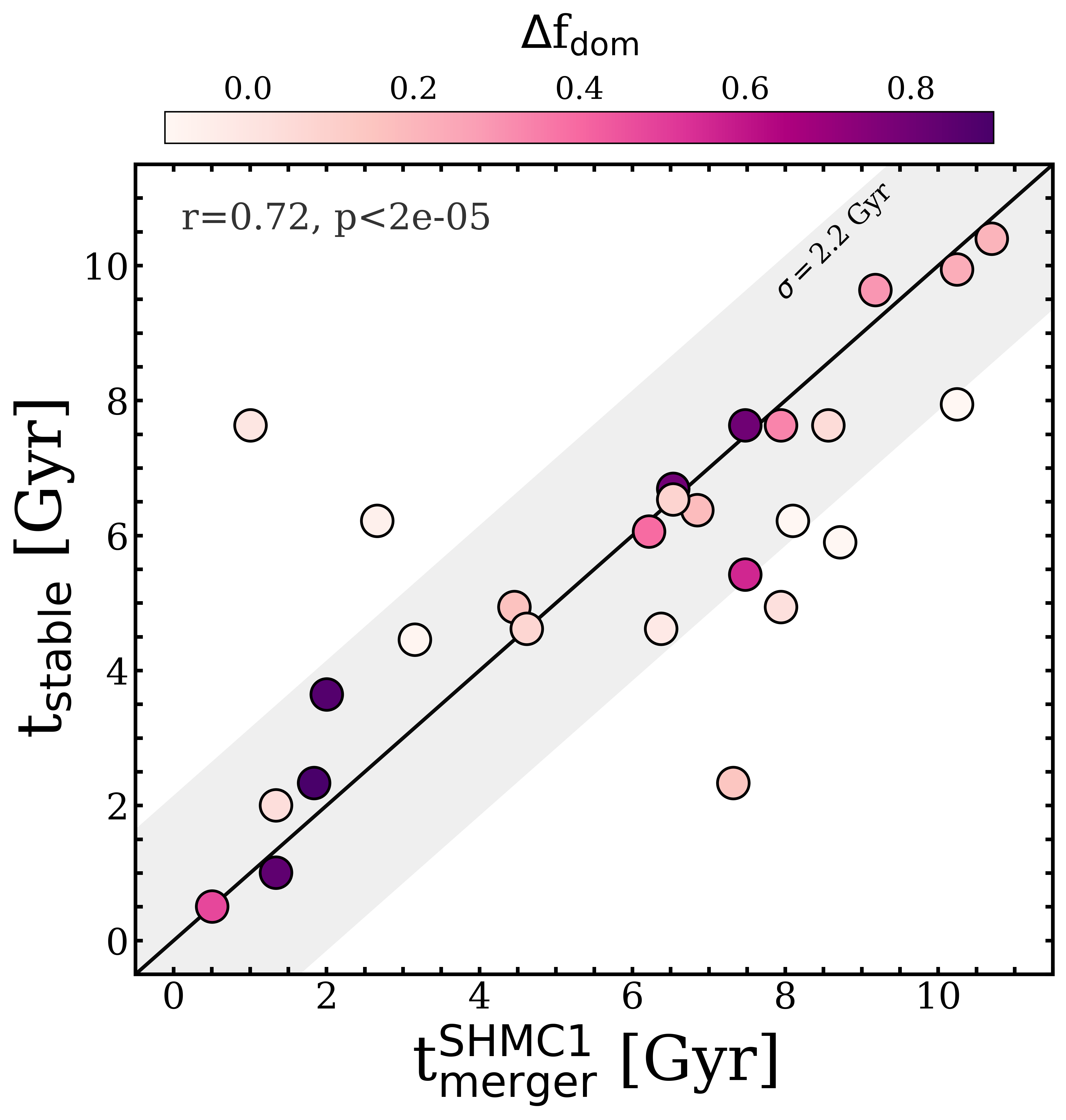}
    \caption{The stability time ($\tst$) as a function of the merger time of the SHMC1 ($\tdisr$), color-coded by the difference between the mass fraction of the SHMC1 and the combined contribution from the SHMC2 and SHMC3 (i.e. $\Delta \rm f_{dom} = f^{SHMC1}_{mass} - f^{SHMC2+SHMC3}_{mass}$). The black line is 1:1 line and the gray area is the standard deviation of the residual between $\tst$ and $\tdisr$.}
    \label{fig:corr_tst_tdis}
\end{figure}

{\renewcommand{\arraystretch}{1.1}
\begin{table*}[]
\caption{Observed properties of nearby galaxies available in the literature and $\tst$ estimated from this article.} 
\centering
    \begin{tabular}{c c c c c c }
    \\ \hline \hline
Galaxy & $\mgal$  & $\tninety$ & $\tst (\mgal)$  & $\tst (\tninety)$ &  References \\
 & $[10^{10}~\msun]$  & [Gyr] & [Gyr] & [Gyr] & $\mgal$, $\tninety$ \\ \hline
NGC 253 & 5.5 $\pm$ 1.4   & 6.2 $\pm$ 1.5  & 7.8 & 5.0 &  H17, H23 \\
NGC 891 & 5.3 $\pm$ 1.3  & 6.9 $\pm$ 1.6  & 7.8 & 5.6 &  H17, H23 \\
NGC 3031 & 5.6 $\pm$ 1.4 & 11.8 $\pm$ 4.0 & 9 & 9.9 &  H23, H23  \\
NGC 5128 & 20  & 2.5 $\pm$ 0.5  & 9.1 & 1.7  & F\&R18, H17 \\
NGC 4565 & 8.0 $\pm$ 2.0   &  & 8.2  &   & H17  \\
NGC 4945 & 3.8 $\pm$ 0.95  &  & 7.4  &   & H17   \\
NGC 7814 & 4.5 $\pm$ 1.1   &  & 7.6  &   & H17    \\
M31 & 10.3 $\pm$ 2.3  & 2.5 $\pm$ 0.5  &  8.4  & 1.7  & L\&N15, H23 \\
MW & 6.1 $\pm$ 1.1   & 10 & 7.9 & 7.5  & S15, H23  \\ \hline \\
\end{tabular}
\tablefoot{From left to right: Galaxy, name of the observed galaxy; $\mgal$, observed stellar galaxy mass; $\tninety$, observational time at which 90 per cent of the stars are formed; $\tst$ lookback time estimated from the linear regression with galaxy stellar mass (see Sec.~\ref{sec:observations}); $\tst$ lookback time inferred from the correlation with $\tninety$ following eq.~\ref{eq:tst:a}. References from which the $\mgal$ and $\tninety$ were obtained, H17: \cite{Harmsen_2017}, H23: \cite{Harmsen_2023}, F\&R18: \citet{Fall_2018}, L\&N15: \citet{Licquia_2015}, S15: \citet{Sick_2015}. Estimations of $\tst$ using $\tninety$ have an uncertainty of 2.8~Gyr calculated by error propagation of the standard deviations. Estimations of $\tst$ based on $\mgal$ have an uncertainty associated to the standard deviation of the relation shown in Fig.~\ref{fig:tst_Mgal} ($\sigma=2.6$~Gyr).}
\label{table:Galaxies}
\end{table*}}

\section{Observational implications}\label{sec:observations}

Our findings suggest that an estimation of the stability time provides a proxy of the merger time of the main contributing satellite that build up the stellar halo of galaxies, and combined with an estimation of $\tdisr$ could provide clues about the relative importance of the second and third main progenitors. We explore possible observational proxies to infer the $\tst$.

Recent observational advances have made it possible to estimate the stellar halo age of nearby galaxies using GHOSTS data. \citet{Harmsen_2023} showed that the AGB/RGB ratio can be used to estimate when star formation shuts off in a halo, with an uncertainty of at least $\pm2$~Gyr. By comparing the AGB/RBG ratio with stellar population models and inferred star formation histories for observed dwarf galaxies, in \citet{Harmsen_2023} were able to constrain $\tninety$, defined as the time before which 90 percent of the halo stars were formed. In order to mimic this observational estimation, we computed $\tninety$ from the cumulative stellar mass as a function of age of the stars belonging to the simulated stellar halo at $z=0$, and we will explore its connection with $\tst$. Additionally, we built the halo cardiogram using the age-metallicity relationship (AMR) of the simulated halo stars at $z=0$. In this way, we also estimate a proxy for the stability time ($\tamr$) from the age-metallicity relation, defined as the age of the populations at which their cumulative metallicity reach a value 0.1~dex of the overall halo metallicity at $z=0$. In what follows, we explore the relation between $\tst$ and both $\tninety$ and $\tamr$.

\begin{figure}
    \centering
    \includegraphics[width=0.95\linewidth]{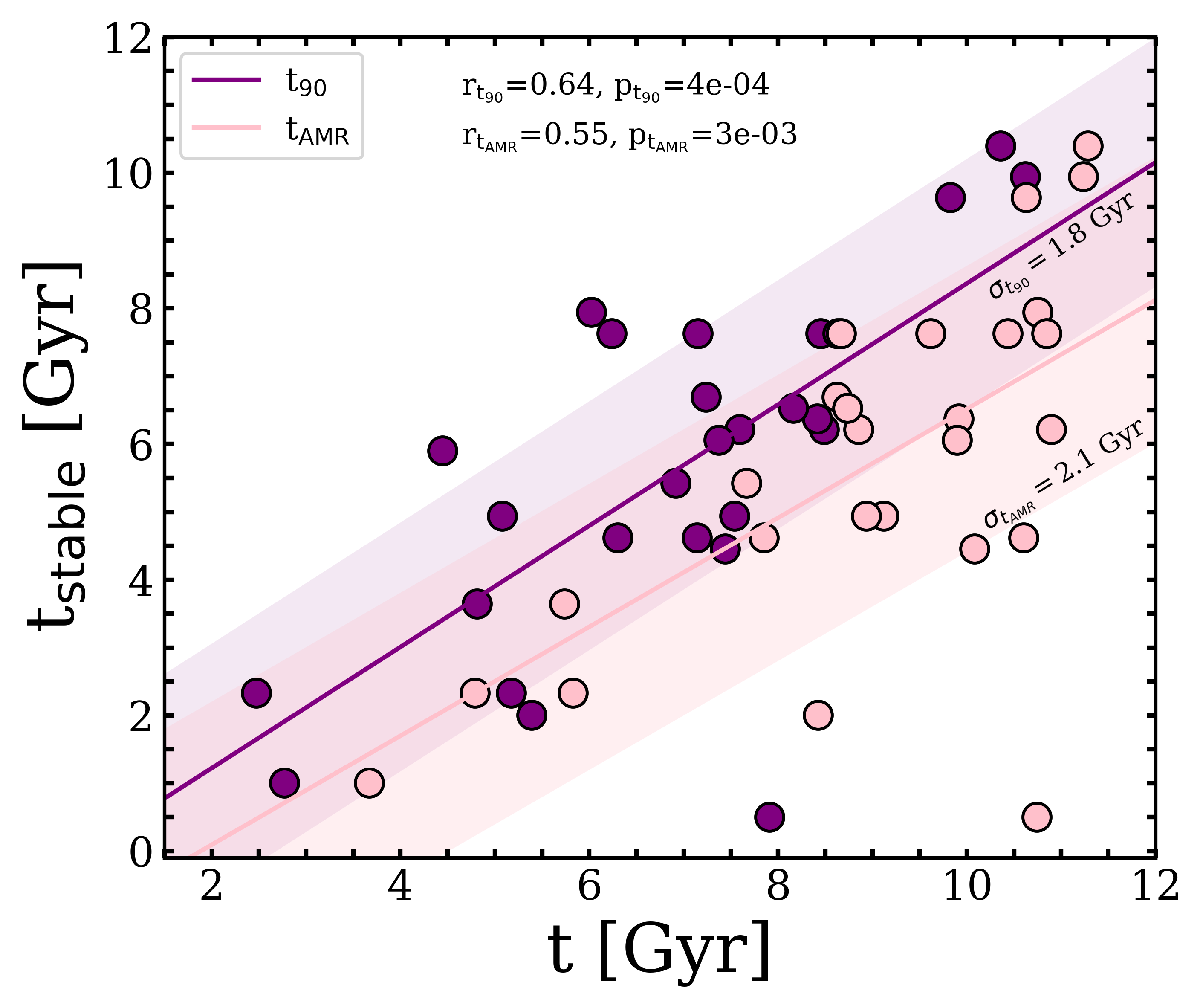}
    \caption{The stability time, $\tst$, as function of the time: $\tninety$ (purple) defined as the time at which 90 percent of the stars formed and $\tamr$ (pink) defined as an observed stability time from the age-metallicity at $z=0$. Colored lines are linear fits and shaded regions denote the standard deviation of the data. Spearman and p-value coefficient are display at the top center.}
    \label{fig:tst_t90}
\end{figure}

Figure~\ref{fig:tst_t90} shows the correlation between $\tst$ and $\tninety$ (purple circles, $\rm r = 0.64$ and p<4e-04), and $\tamr$ (pink circles, $\rm r = 0.55$ and $\rm p<3e-03$). Both relations can be fitted by linear regressions,  

\begin{subequations}\label{eq:tst}
\begin{align}
  \tst &= 0.89\, \tninety - 0.56 \label{eq:tst:a}\\
  \tst &= 0.80\, \tamr   - 1.51 \label{eq:tst:b}
\end{align}
\end{subequations}

The scatter is quantified by the standard deviation of $\sigma=1.8$ and 2.1~Gyr for $\tst$-$\tninety$ and $\tst$-$\tamr$, respectively.  
These results suggest that the epoch at which the stellar halo metallicity stabilizes is closely linked to the time by which most of the halo stars were formed, $\tninety$. In this sense, the stability time can be interpreted as a chemical proxy for the halo formation time. It is important to note that halo stars mostly form in dwarf galaxies at early epochs and are accreted into the host galaxy at later times. For this reason, we emphasize the distinction between $\tninety$ and the time at which those stars were actually accreted into the stellar halo. 

Hence, we use $\tninety$ measurements from GHOSTS data \citep{Harmsen_2023} and our predicted relation given by equation~\ref{eq:tst:a} to estimate $\tst$. These predictions are displayed in Table~\ref{table:Galaxies} and the implications are discussed in the next Section. We do not report estimations of $\tamr$ for these galaxies because there are not resolved stellar populations available. An estimation could be attempted for the MW but it implies a major observational work which is beyond the scope of this paper. 
 
Another relevant observable is the stellar mass of galaxies. We find a trend between the galaxy stellar mass and $\tst$, where more massive galaxies tend to stabilize their metallicity at earlier times than low-mass galaxies, consistent with a downsizing scenario. The relation is displayed in Fig.~\ref{fig:tst_Mgal}. The scatter ($\sigma=2.6$~Gyr) is significant as expected on the basis of the previous discussion where we show that each halo has its own evolutionary track even more for similar masses (Fig.~\ref{fig:example_paths}).
Although, the relation is clear with $\rm r = 0.39$ ($\rm p = 0.05$), it remains weak. Hence, it should be interpreted with caution due to the large variability (see Sec.~\ref{sec:discussion} for further discussion).
A linear regression fit yields $\tst = 2.25\times log_{10}\ \mgal -16.33$).

We stress the fact that Fig.~\ref{fig:tst_Mgal} shows a relation between $\tst$ and galaxy stellar mass. Previous works such as \citet{Harmsen_2023} have reported an anti-correlation between $\tninety$ and the accreted stellar halo mass. A detailed discussion on this point can be found in Appendix~\ref{appendix:SH-t90}. However, given the large scatter in this relation, which was reported by \citet{Harmsen_2023} using TNG50 at the low-mass end, we do not carry out a more detailed comparison in the present work.

\begin{figure}
    \centering
    \includegraphics[width=0.95\linewidth]{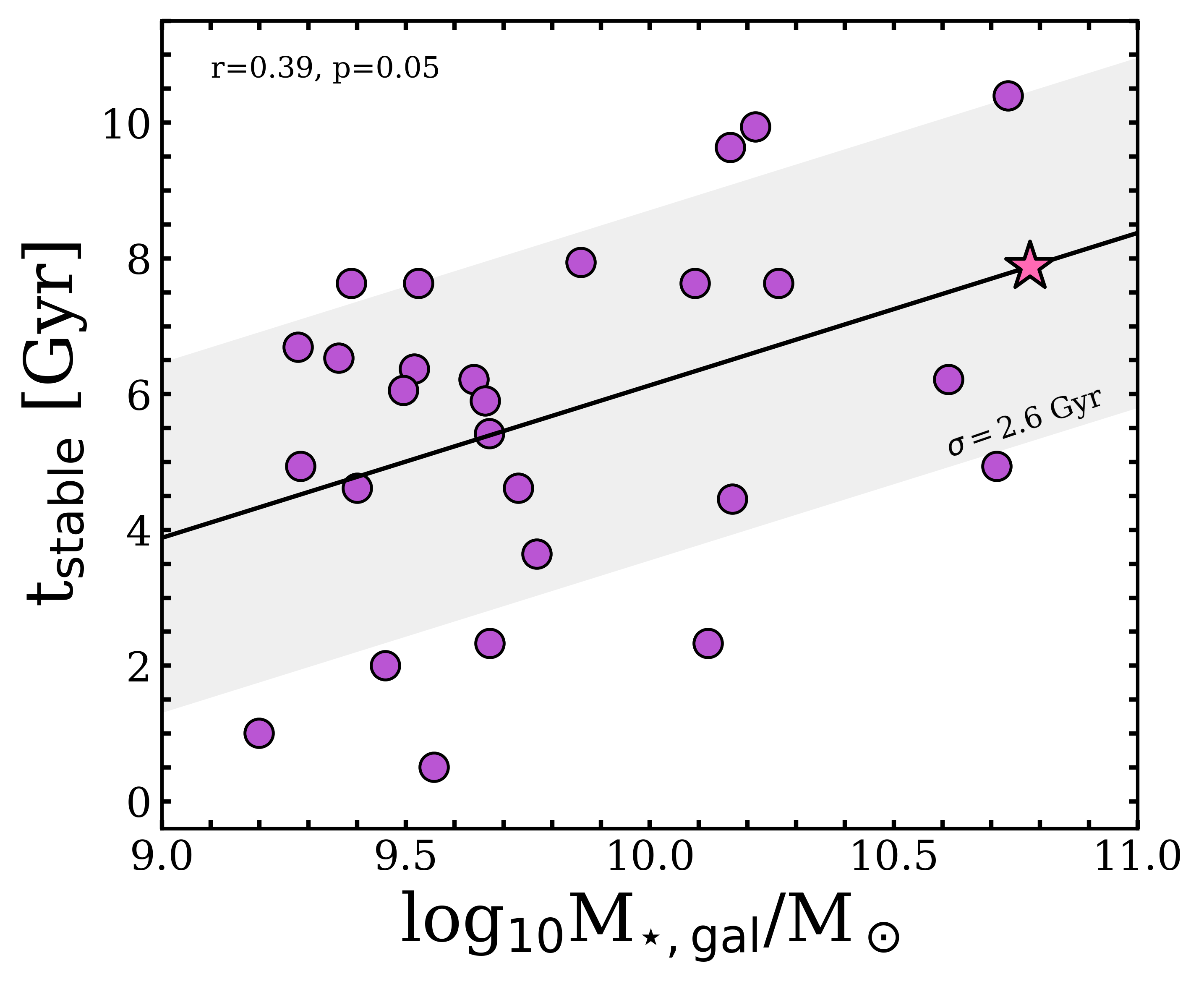}
    \caption{Stability time, $\tst$, as a function of stellar galaxy mass, $\rm M_{\star,\rm gal}$. The black line corresponds to a linear regression shown in eq.~\ref{eq:tst:a}, and the gray shaded region denotes the standard deviation of $\tst$. The Milky Way is included as a reference, adopting $\mgal = 6.08 \pm 1.14 \times 10^{10}\ \msun$ from \citep{Sick_2015} (pink star).}
    \label{fig:tst_Mgal}
\end{figure}

\section{Discussion}\label{sec:discussion}

Our results reveal a correlation between $\tdisr$ and $\tst$ that follows a one-to-one relation, with a scatter of $2.2$~Gyr (Fig.~\ref{fig:corr_tst_tdis}). This correlation arises because SHMC1 is typically more massive, and following the MZ$_\star$R, more metal-rich than other contributing satellites, which tend to be less massive  \citep[see also][]{Munoz-escobar_2025}. Consequently, for halos in the one-to-one line in the $\tst-\tdisr$, the debris of SHMC1 sets the halo metallicity at $\tst$, and later accretions do not significantly alter the median metallicity established at $\tst$. This is consistent with the results of \citet{Harmsen_2017}, who suggest that the main contributor sets the metallicity of the halo, and the following numerical works by \citet{DSouza_2018} and \citet{Monachesi_2019}. Halos with larger residuals in the $\tdisr-\tst$ plane, have SHMC2 and SHMC3 that together contribute with comparable or higher mass fractions than  SHMC1 (i.e. smaller $\Delta \rm f_{dom}$), implying that the median halo metallicity is not set by the SHMC1, and other satellite contributors are needed. This occurs because, although the stellar mass of the satellite is important in determining its metallicity, the mass contributed by the satellite to the stellar halo also depends on its orbital parameters, which will influence how the material is distributed in the halo and galaxy. Furthermore, whether the satellite is fully disrupted or survives longer in the halo, or if it is gas-rich or gas-poor also influence their metallicity and the metal content that contributed to the halo, since they could  sustain star formation longer, increasing their metallicity.

By using our predicted relations (eq.~\ref{eq:tst:a}), we estimate that the last major merger of the MW took place at $\sim 7.5$~Gyr, consistent, within uncertainties, with the $\sim 8-11$~Gyr window associated with Gaia Enceladus Sausage (GES) \citep{Helmi_2018,Belokurov_2018, Di_Matteo_2019,Gallart_2019, Giribaldi_2023}. If GES is the last major accretion event of the MW that built up the bulk of the stellar halo, its debris would drive the present stellar halo metallicity. However, if other accretion events such as Sagittarius, contribute significantly to the halo assembly, the merger time obtained from our predictions would be slightly underestimated. 

Table~\ref{table:Galaxies} also provides predictions for NGC 3031 (M81). This galaxy is in an early stage interaction with M82 and NGC 3077 \citep{Smercina_2020}. \citet{Harmsen_2023} derived $\tninety=11.8\pm4$, noting that the AGB/RGB ratio is poorly measured, which makes this estimation of $\tninety$ uncertain. An independent constraint comes from \citet{Durrell_2010} who report a mean age\footnote{Using deep color-magnitude diagram from the Hubble Space Telescope, they obtained the mean age from the shape of the red giant branch, the magnitude of the red clump, and the location of the red giant branch bump.} of $9\pm2$~Gyr for the dominant halo population of a field at a projected distance of 19~kpc from the center of M81. Using that mean age in our predicted relation (eq.~\ref{eq:tst:a}), we infer $\tst=7.81$~Gyr, which is $\sim2$~Gyr younger than the $\tst\approx10$~Gyr inferred from the $\tninety$ measurement of \citet{Harmsen_2023}. If the stellar halo of M81 is dominated by debris of a single merger, we predict that this event happened in the early stage of M81. However, because of the large merger rate (1:2) estimated for the ongoing interaction with M82, we expect that the stellar halo of M81 might be temporarily displaced from the MZhR until the major accretion event ceases.  An approach to estimate the scatter in the $\tst$-$\tdisr$ plane would be to estimate the relative contribution coming from M82.

For Andromeda galaxy, M31, the $\mgal-\tst$ and $\tninety-\tst$ relations yield markedly different merger times (assuming $\tst \approx \tdisr$), suggesting that M31 may represent an outlier in Fig.~\ref{fig:corr_tst_tdis}. One possible interpretation is that its stellar halo has been influenced by a recent accretion event, bringing a substantial contribution of stellar material to the halo. 
From $\mgal$-$\tst$ relation, we infer the last major merger took place about $8.4$~Gyr ago. In contrast, the $\tninety$-$\tst$ relation gives an estimation of the merger time of $1.7$~Gyr ago. The latter one agrees with evidence of an active recent merger history suggested by a $\sim2-3$~Gyr burst of star formation that can be explained by a recent major merger event that triggered it \citep{Hammer_2018, Dsouza_M31_2018, Escala_2020, Tsakonas_2025}. It is important to note that the estimate of $\tninety$ for M31 by \citet{Harmsen_2023} was not performed using the AGB/RGB ratio, instead it is based on evidence indicating that the last significant episode of inner stellar halo star formation occurred approximately 2–3 Gyr ago. Therefore, future observations are need to consistently measure the $\tninety$ of M31. Finally, our findings suggest that M31 just reached the MZhR at $z=0$ consistent with the observational data shown in Fig.~\ref{sec:mzhr_z0}. Moreover, if its last major merger contributed with high $\Delta \rm f_{dom}$, its metallicity would be driven by that event, which we predict happened at about $1.7$~Gyr ago. 

Even though M31 and MW have comparable stellar galaxy masses, they show different accretion histories which lead to different stability times as well as distinct stellar halo masses and metallicities. Hence they are expected to trace different evolutionary paths as  shown in the left column in Fig.~\ref{fig:example_paths}.

Finally, the stability time inferred from the galaxy stellar mass relation should be interpreted with caution due to the diversity in formation histories of galaxies, reflected in the variety of halo mass fractions at a fixed $\mgal$, as was discussed for M31 and MW. The large scatter ($\sigma=2.6$~Gyr) in the $\mgal$–$\tst$ plane might reflects the stochastic nature of galaxy formation and limits the precision of $\tst$ inferred from $\mgal$ alone. Larger samples are needed to test this trend and probe any secondary dependence.

A numerical caveat to our result is the uncertainty in defining the moment when stellar material from a dwarf galaxy becomes part of the main halo (referred to as merger time in this work). This timing likely depends on several factors, including  the subhalo identification algorithm and the merger tree construction method. Observationally, identifying this transition is equally challenging, as it is not always clear when a tidally stripped dwarf galaxy should no longer be classified as a distinct system (e.g., the case of the Sagittarius dwarf in the Milky Way). Additionally, stellar halos are expected to have in-situ stars, which might have different origin. \citetalias{Gonzalez-Jara_2025} showed that for these outer halos the fraction of in-situ is about 20\%. However, we note that in our estimations, we have considered them to calculate $\tninety$ and hence, the simulated relation between $\tst$ and $\tninety$ already considered the possible presence of in-situ stars at higher galactocentric distances. Finally, we note that the estimated values of $\tninety$ depend on the adopted stellar population (see Appendix~\ref{appendix:SH-t90}). However, regardless of whether ex-situ, accreted, or all stellar particles are considered, we consistently find a correlation between $\tninety$ and $\tst$, although the slope of the relation varies. This highlights the sensitivity of halo assembly times to the definition of the stellar component and suggests that such relations can provide a useful test of the subgrid physics implemented in simulations. A similar dependence is also found in observations, \citet{Harmsen_2023} show that the inferred $\tninety$ varies depending on the halo region used in their analysis of individual galaxies, reflecting the complexity of stellar halos and the expected presence of substructures. These results emphasize the importance of assessing which stellar populations are used to study the properties of the stellar halos.
 
\section{Summary and conclusions}\label{sec:summary}

We used \cielo~zoom-in simulations \citep{Tissera_2025} to investigate the stellar MhZR of galaxies with stellar masses in the range of log$_{10}(\mgal/\msun)\in[9, 11]$. Stellar halos were isolated using the AM-E method \citep{Tissera_2012} following the procedure in \citetalias{Gonzalez-Jara_2025}. We identified the stellar halo between 1.5r$_{\rm opt}$ and r$_{\rm rvir}$, from z$\approx3.5$ to $z=0$, to evaluate the existence of the MZhR at high redshift and characterize whether it evolves with time.

The \cielo~galaxies reproduce the MZhR at $z=0$, in agreement with previous observational and theoretical studies \citep{DSouza_2018, Monachesi_2019, Smercina_2022, Tau_2025}. We propose to consider an aperture to measure the halo metallicity that scale with the size of the galaxy, instead of considering the 30~kpc usually used for MW-like galaxies. The correlation between stellar halo mass and metallicity in CIELO holds whether we consider an aperture comparable to that used in GHOSTS observations or the stellar halo within 1.5$\ropt$ to $\rvir$ (Fig~\ref{fig:MhZR_z0}). In both cases in-situ stars were considered, as their contribution is less than $\sim$15\% and 20\%, respectively, and observational estimates typically include all stellar populations due to the difficulty of isolating accreted stars. We note that our sample spans a broader range of galaxy morphologies and masses, whereas previous studies focused primarily on Milky Way mass-like galaxies.

We find a well-defined MZhR from $z\approx 3.5$ to $z=0$, described by linear fits, that slightly evolves  while preserving the slope at $z=0$ (Fig.~\ref{sec:MZ_evolution}). For reference, a stellar halo with $\mhalo = 10^{9}~\msun$ increases its metallicity by $\sim 0.21$~dex from $z\approx3.5$ to $z=0$. These results are consistent with a progressive accretion of dwarf galaxies, which follow a mass-metallicity relation, driving the systematic enrichment shown in the stellar halos over time. 

To characterize when halos settle on the present-day MZhR, we define the halo cardiograms (median halo metallicity relative to their values at $z=0$ as a function of time, Fig.~\ref{fig:example_cardiogram}) to identify the lookback time ($\tst$) at which the median stellar halo metallicity remains within $\pm 0.1$ dex of its value at $z=0$. We find that $\tst$ and the merger time of the first stellar halo main contributor (SHMC1) follows a one-to-one relation, with a scatter of $\sim2$~Gyr (Fig.~\ref{fig:corr_tst_tdis}). When the mass fraction of SHMC1 is higher than the remaining contributors by at least 20\%, the stellar halo metallicity is stabilized once SHMC1 is fully disrupted. In this case, the stability time can be used as a proxy to estimate the merger time of the main contributing satellite that build up the stellar halo. This relation has a dispersion due to the fact that for some halos, the second and third contributors are significant and could contribute to set distinct evolutionary paths. 

We show that an observational cardiogram of the stellar halos can be built from the age-metallicity relation. By applying the same criteria used for the simulated cardiograms, we estimated $\tamr$, which is found to correlate with $\tst$. However, this is currently difficult to perform observational due to low surface brightness of stellar halos. Hence, we explore $\tninety$ as an observational proxy for dating the moment at which a stellar halo reaches the local MZhR and, by using the simulated correlation shown in eq.~\ref{eq:tst:a}, estimating the time of the main merger event that contributed to built up it. We find an excellent correlation between $\tst$ and $\tninety$, which we use to predict the time of complete disruption of the main contributor of several observed halos including the MW (Table~\ref{table:Galaxies}). For the MW, our estimate yields $\sim 7.5 \pm 2.8$~Gyr, with the uncertainty estimated via error propagation of the standard deviation of eq.\ref{eq:tst:a} and $\tst \approx \tninety$, consistent within the uncertainties with the $\sim8-10$~Gyr window associated with GES, although this estimate would be less robust if the Sagittarius merger contributed a larger mass fraction to the MW’s stellar halo. If we take 10 Gyr as the reference time for the GES merger and consider it as SHMC1, then our findings suggest that Sagittarius contributes significantly to the build up of the MW's stellar halo. On the other hand, for M31, we predict a recent stability scenario for the MZhR, delayed by a merger event that we predict took place at $\sim1.7$~Gyr ago. 

Future estimations of the stellar metallicities of galaxies and their stellar halos at low and high redshift will provide crucial data for test our predictions and improve our understanding of galaxy formation.

\begin{acknowledgements}
    We thank the anonymous referee for the valuable comments that helped improve this paper. We are also grateful for useful comments by Paula Jofré, Claudia Lagos and Rolando Dunner. JGJ acknowledges funding by ANID (Beca Doctorado Nacional, Folio 21210846). JGJ, PBT and AM acknowledge support from the ANID BASAL project FB210003. PBT acknowledges partial funding by Fondecyt-ANID 1240465/2024. AM acknowledges support from the ANID FONDECYT Regular grant 1251882. BTC gratefully acknowledges funding by ANID (Beca Doctorado Nacional, Folio 21232155). SP acknowledge partial support from CONICET through grant PIP 11220210100214. CCV thanks comite mixto ESO-Chile 2024. RDT thanks the Ministerio de Ciencia e Innovación (Spain) for financial support under Project grant PID2021-122603NB-C21 as well as Project PID2024-156100NB-C21 financed by MICIU/AEI/10.13039/501100011033/FEDER, EU. This work has received funding from the European Union’s HORIZON-MSCA-2021-SE-01 Research and Innovation Programme under the Marie Sklodowska-Curie grant agreement number 101086388 - Project acronym:LACEGAL. This project used the Ladgerda Cluster (Fondecyt 1200703/2020 hosted at the Institute for Astrophysics, Chile), the NLHPC (Centro de Modelamiento Matemático, Chile), Geryon clusters (Center for Astrophysics, CATA, Chile), and the Barcelona Supercomputer Center (Spain).

\end{acknowledgements}

\bibliographystyle{aa} 
\bibliography{bibliography.bib}

@ARTICLE{miranda2025,
       author = {{Miranda}, Valentina P. and {Tissera}, Patricia B. and {Sillero}, Emanuel and {Gonzalez-Jara}, Jenny and {Bignone}, Lucas and {Mu{\~n}oz-Escobar}, Ignacio and {Pedrosa}, Susana and {Dom{\'\i}nguez-Tenreiro}, Rosa},
        title = "{Metal-loaded outflows in sub-Milky Way galaxies in the CIELO simulations}",
      journal = {\aap},
         year = 2026,
        month = feb,
       volume = {706},
          eid = {A274},
        pages = {A274},
          doi = {10.1051/0004-6361/202556691},
archivePrefix = {arXiv},
       eprint = {2511.19630},
 primaryClass = {astro-ph.GA},
       adsurl = {https://ui.adsabs.harvard.edu/abs/2026A&A...706A.274M}
}

@article{Krikler_1987,
title = {Historical Aspects of Electrocardiography},
journal = {Cardiology Clinics},
volume = {5},
number = {3},
pages = {349-355},
year = {1987},
note = {12-Lead Electrocardiography},
issn = {0733-8651},
doi = {https://doi.org/10.1016/S0733-8651(18)30525-3},
url = {https://www.sciencedirect.com/science/article/pii/S0733865118305253},
author = {Dennis M. Krikler}
}

@article{Fye_1994,
  author       = {Fye, W.~B.},
  title        = {A history of the origin, evolution, and impact of electrocardiography},
  journal      = {Am. J. Cardiol.},
  volume       = {73},
  number       = {13},
  pages        = {937--949},
  year         = {1994},
  doi          = {10.1016/0002-9149(94)90135-x}
}

@article{Gonzalez-Jara_2025,
	author = {{Gonzalez-Jara}, Jenny and {Tissera, Patricia B.} and {Monachesi, Antonela} and {Sillero, Emanuel} and {Pallero, Diego} and {Pedrosa, Susana} and {Tau, Elisa A.} and {Tapia-Contreras, Brian} and {Bignone, Lucas}},
	title = {Unveiling the formation channels of stellar halos through their chemical fingerprints},
	DOI= "10.1051/0004-6361/202452639",
	url= "https://doi.org/10.1051/0004-6361/202452639",
	journal = {A\&A},
	year = 2025,
	volume = 693,
	pages = "A282",
}

@ARTICLE{Tau_2025,
       author = {{Tau}, Elisa A. and {Monachesi}, Antonela and {G{\'o}mez}, Facundo A. and {Grand}, Robert J.~J. and {Pakmor}, R{\"u}diger and {van de Voort}, Freeke and {Marinacci}, Federico and {Bieri}, Rebekka},
        title = "{Age and metallicity of low-mass galaxies: from their centres to their stellar halos}",
      journal = {arXiv e-prints},
         year = 2025,
        month = nov,
          eid = {arXiv:2511.20806},
        pages = {arXiv:2511.20806},
          doi = {10.48550/arXiv.2511.20806},
archivePrefix = {arXiv},
       eprint = {2511.20806},
 primaryClass = {astro-ph.GA},
       adsurl = {https://ui.adsabs.harvard.edu/abs/2025arXiv251120806T}
}

@ARTICLE{Tau_ageFeH_2025,
       author = {{Tau}, Elisa A. and {Monachesi}, Antonela and {G{\'o}mez}, Facundo A. and {Grand}, Robert J.~J. and {Pakmor}, R{\"u}diger and {van de Voort}, Freeke and {Marinacci}, Federico and {Bieri}, Rebekka},
        title = "{Age and metallicity of low-mass galaxies: from their centres to their stellar halos}",
      journal = {arXiv e-prints},
         year = 2025,
        month = nov,
          eid = {arXiv:2511.20806},
        pages = {arXiv:2511.20806},
          doi = {10.48550/arXiv.2511.20806},
archivePrefix = {arXiv},
       eprint = {2511.20806},
 primaryClass = {astro-ph.GA},
       adsurl = {https://ui.adsabs.harvard.edu/abs/2025arXiv251120806T}
}

@ARTICLE{Gozman_2023,
       author = {{Gozman}, Katya and {Bell}, Eric F. and {Smercina}, Adam and {Price}, Paul and {Bailin}, Jeremy and {de Jong}, Roelof S. and {D'Souza}, Richard and {Jang}, In Sung and {Monachesi}, Antonela and {Slater}, Colin},
        title = "{Saying Hallo to M94's Stellar Halo: Investigating the Accretion History of the Largest Pseudobulge Host in the Local Universe}",
      journal = {\apj},
         year = 2023,
        month = apr,
       volume = {947},
       number = {1},
          eid = {21},
        pages = {21},
          doi = {10.3847/1538-4357/acbe3a},
archivePrefix = {arXiv},
       eprint = {2304.08436},}

@ARTICLE{Panter_2008,
       author = {{Panter}, Benjamin and {Jimenez}, Raul and {Heavens}, Alan F. and {Charlot}, Stephane},
        title = "{The cosmic evolution of metallicity from the SDSS fossil record}",
      journal = {\mnras},
         year = 2008,
        month = dec,
       volume = {391},
       number = {3},
        pages = {1117-1126},
          doi = {10.1111/j.1365-2966.2008.13981.x},
archivePrefix = {arXiv},
       eprint = {0804.3091},
 primaryClass = {astro-ph},
       adsurl = {https://ui.adsabs.harvard.edu/abs/2008MNRAS.391.1117P}
}

@ARTICLE{Murphy_2022,
       author = {{Murphy}, Geoff G. and {Yates}, Robert M. and {Mohamed}, Shazrene S.},
        title = "{L-GALAXIES 2020: the formation and chemical evolution of stellar haloes in Milky Way analogues and galaxy clusters}",
      journal = {\mnras},
         year = 2022,
        month = feb,
       volume = {510},
       number = {2},
        pages = {1945-1963},
          doi = {10.1093/mnras/stab3568},
archivePrefix = {arXiv},
       eprint = {2111.12737},
 primaryClass = {astro-ph.GA},
       adsurl = {https://ui.adsabs.harvard.edu/abs/2022MNRAS.510.1945M}
}

@ARTICLE{Ma_2015,
       author = {{Ma}, Xiangcheng and {Hopkins}, Philip F. and {Faucher-Gigu{\`e}re}, Claude-Andr{\'e} and {Zolman}, Nick and {Muratov}, Alexander L. and {Kere{\v{s}}}, Du{\v{s}}an and {Quataert}, Eliot},
        title = "{The origin and evolution of the galaxy mass-metallicity relation}",
      journal = {\mnras},
         year = 2016,
        month = feb,
       volume = {456},
       number = {2},
        pages = {2140-2156},
          doi = {10.1093/mnras/stv2659},
archivePrefix = {arXiv},
       eprint = {1504.02097},
 primaryClass = {astro-ph.GA},
       adsurl = {https://ui.adsabs.harvard.edu/abs/2016MNRAS.456.2140M}
}

@ARTICLE{Langeroodi_2023,
       author = {{Langeroodi}, Danial and {Hjorth}, Jens and {Chen}, Wenlei and {Kelly}, Patrick L. and {Williams}, Hayley and {Lin}, Yu-Heng and {Scarlata}, Claudia and {Zitrin}, Adi and {Broadhurst}, Tom and {Diego}, Jose M. and {Huang}, Xiaosheng and {Filippenko}, Alexei V. and {Foley}, Ryan J. and {Jha}, Saurabh and {Koekemoer}, Anton M. and {Oguri}, Masamune and {Perez-Fournon}, Ismael and {Pierel}, Justin and {Poidevin}, Frederick and {Strolger}, Lou},
        title = "{Evolution of the Mass-Metallicity Relation from Redshift z {\ensuremath{\approx}} 8 to the Local Universe}",
      journal = {\apj},
         year = 2023,
        month = nov,
       volume = {957},
       number = {1},
          eid = {39},
        pages = {39},
          doi = {10.3847/1538-4357/acdbc1},
archivePrefix = {arXiv},
       eprint = {2212.02491},
 primaryClass = {astro-ph.GA},
       adsurl = {https://ui.adsabs.harvard.edu/abs/2023ApJ...957...39L}
}

@ARTICLE{Dominguez_Gomez_2023,
       author = {{Dom{\'\i}nguez-G{\'o}mez}, Jes{\'u}s and {P{\'e}rez}, Isabel and {Ruiz-Lara}, Tom{\'a}s and {Peletier}, Reynier F. and {S{\'a}nchez-Bl{\'a}zquez}, Patricia and {Lisenfeld}, Ute and {Bidaran}, Bahar and {Falc{\'o}n-Barroso}, Jes{\'u}s and {Alc{\'a}zar-Laynez}, Manuel and {Argudo-Fern{\'a}ndez}, Mar{\'\i}a and {Bl{\'a}zquez-Calero}, Guillermo and {Courtois}, H{\'e}l{\`e}ne and {Duarte Puertas}, Salvador and {Espada}, Daniel and {Florido}, Estrella and {Garc{\'\i}a-Benito}, Rub{\'e}n and {Jim{\'e}nez}, Andoni and {Kreckel}, Kathryn and {Rela{\~n}o}, M{\'o}nica and {S{\'a}nchez-Menguiano}, Laura and {van der Hulst}, Thijs and {van de Weygaert}, Rien and {Verley}, Simon and {Zurita}, Almudena},
        title = "{Stellar mass-metallicity relation throughout the large-scale structure of the Universe: CAVITY mother sample}",
      journal = {\aap},
         year = 2023,
        month = dec,
       volume = {680},
          eid = {A111},
        pages = {A111},
          doi = {10.1051/0004-6361/202346884},
archivePrefix = {arXiv},
       eprint = {2310.11412},
 primaryClass = {astro-ph.GA},
       adsurl = {https://ui.adsabs.harvard.edu/abs/2023A&A...680A.111D}
}

@ARTICLE{Torrey_2019,
       author = {{Torrey}, Paul and {Vogelsberger}, Mark and {Marinacci}, Federico and {Pakmor}, R{\"u}diger and {Springel}, Volker and {Nelson}, Dylan and {Naiman}, Jill and {Pillepich}, Annalisa and {Genel}, Shy and {Weinberger}, Rainer and {Hernquist}, Lars},
        title = "{The evolution of the mass-metallicity relation and its scatter in IllustrisTNG}",
      journal = {\mnras},
         year = 2019,
        month = apr,
       volume = {484},
       number = {4},
        pages = {5587-5607},
          doi = {10.1093/mnras/stz243},
archivePrefix = {arXiv},
       eprint = {1711.05261},
 primaryClass = {astro-ph.GA},
       adsurl = {https://ui.adsabs.harvard.edu/abs/2019MNRAS.484.5587T}
}

@ARTICLE{Naidu_2022,
       author = {{Naidu}, Rohan P. and {Conroy}, Charlie and {Bonaca}, Ana and {Zaritsky}, Dennis and {Ting}, Yuan-Sen and {Caldwell}, Nelson and {Cargile}, Phillip A. and {Speagle}, Joshua S. and {Chandra}, Vedant and {Johnson}, Benjamin D. and {Woody}, Turner and {Han}, Jiwon Jesse},
        title = "{Live Fast, Die $\alpha$-Enhanced: The Mass-Metallicity-$\alpha$ Relation of the Milky Way's Disrupted Dwarf Galaxies}",
      journal = {arXiv e-prints},
         year = 2022,
        month = apr,
          eid = {arXiv:2204.09057},
        pages = {arXiv:2204.09057},
          doi = {10.48550/arXiv.2204.09057},
archivePrefix = {arXiv},
       eprint = {2204.09057},
 primaryClass = {astro-ph.GA},
       adsurl = {https://ui.adsabs.harvard.edu/abs/2022arXiv220409057N}
}

@ARTICLE{Grimozzi_2024,
       author = {{Grimozzi}, Salvador E. and {Font}, Andreea S. and {De Rossi}, Mar{\'\i}a Emilia},
        title = "{Differences in the properties of disrupted and surviving satellites of Milky-Way-mass galaxies in relation to their host accretion histories.}",
      journal = {\mnras},
         year = 2024,
        month = may,
       volume = {530},
       number = {1},
        pages = {95-116},
          doi = {10.1093/mnras/stae878},
archivePrefix = {arXiv},
       eprint = {2401.04182},
 primaryClass = {astro-ph.GA},
       adsurl = {https://ui.adsabs.harvard.edu/abs/2024MNRAS.530...95G}
}

@ARTICLE{Trussler_2019,
       author = {{Trussler}, James and {Maiolino}, Roberto and {Maraston}, Claudia and {Peng}, Yingjie and {Thomas}, Daniel and {Goddard}, Daniel and {Lian}, Jianhui},
        title = "{Both starvation and outflows drive galaxy quenching}",
      journal = {\mnras},
         year = 2020,
        month = feb,
       volume = {491},
       number = {4},
        pages = {5406-5434},
          doi = {10.1093/mnras/stz3286},
archivePrefix = {arXiv},
       eprint = {1811.09283},
 primaryClass = {astro-ph.GA},
       adsurl = {https://ui.adsabs.harvard.edu/abs/2020MNRAS.491.5406T}
}

@ARTICLE{Dave_2017,
       author = {{Dav{\'e}}, Romeel and {Rafieferantsoa}, Mika H. and {Thompson}, Robert J. and {Hopkins}, Philip F.},
        title = "{MUFASA: Galaxy star formation, gas, and metal properties across cosmic time}",
      journal = {\mnras},
         year = 2017,
        month = may,
       volume = {467},
       number = {1},
        pages = {115-132},
          doi = {10.1093/mnras/stx108},
archivePrefix = {arXiv},
       eprint = {1610.01626},
 primaryClass = {astro-ph.GA},
       adsurl = {https://ui.adsabs.harvard.edu/abs/2017MNRAS.467..115D}
}

@ARTICLE{Maiolino_2008,
       author = {{Maiolino}, R. and {Nagao}, T. and {Grazian}, A. and {Cocchia}, F. and {Marconi}, A. and {Mannucci}, F. and {Cimatti}, A. and {Pipino}, A. and {Ballero}, S. and {Calura}, F. and {Chiappini}, C. and {Fontana}, A. and {Granato}, G.~L. and {Matteucci}, F. and {Pastorini}, G. and {Pentericci}, L. and {Risaliti}, G. and {Salvati}, M. and {Silva}, L.},
        title = "{AMAZE. I. The evolution of the mass-metallicity relation at z > 3}",
      journal = {\aap},
         year = 2008,
        month = sep,
       volume = {488},
       number = {2},
        pages = {463-479},
          doi = {10.1051/0004-6361:200809678},
archivePrefix = {arXiv},
       eprint = {0806.2410},
 primaryClass = {astro-ph},
       adsurl = {https://ui.adsabs.harvard.edu/abs/2008A\&A...488..463M}
}

@ARTICLE{Sanders_2021,
       author = {{Sanders}, Ryan L. and {Shapley}, Alice E. and {Jones}, Tucker and {Reddy}, Naveen A. and {Kriek}, Mariska and {Siana}, Brian and {Coil}, Alison L. and {Mobasher}, Bahram and {Shivaei}, Irene and {Dav{\'e}}, Romeel and {Azadi}, Mojegan and {Price}, Sedona H. and {Leung}, Gene and {Freeman}, William R. and {Fetherolf}, Tara and {de Groot}, Laura and {Zick}, Tom and {Barro}, Guillermo},
        title = "{The MOSDEF Survey: The Evolution of the Mass-Metallicity Relation from z = 0 to z 3.3}",
      journal = {\apj},
         year = 2021,
        month = jun,
       volume = {914},
       number = {1},
          eid = {19},
        pages = {19},
          doi = {10.3847/1538-4357/abf4c1},
archivePrefix = {arXiv},
       eprint = {2009.07292},
 primaryClass = {astro-ph.GA},
       adsurl = {https://ui.adsabs.harvard.edu/abs/2021ApJ...914...19S}
}

@ARTICLE{De_Rossi_2015,
       author = {{De Rossi}, M.~E. and {Theuns}, T. and {Font}, A.~S. and {McCarthy}, I.~G.},
        title = "{The evolution of galaxy metallicity scaling relations in cosmological hydrodynamical simulations}",
      journal = {\mnras},
         year = 2015,
        month = sep,
       volume = {452},
       number = {1},
        pages = {486-501},
          doi = {10.1093/mnras/stv1287},
archivePrefix = {arXiv},
       eprint = {1506.02772},
 primaryClass = {astro-ph.GA},
       adsurl = {https://ui.adsabs.harvard.edu/abs/2015MNRAS.452..486D}
}

@ARTICLE{williams_2024,
       author = {{Williams}, Devin J. and {Damjanov}, Ivana and {Sawicki}, Marcin and {Souchereau}, Harrison and {Chen}, Lingjian and {Desprez}, Guillaume and {George}, Angelo and {Annunziatella}, Marianna and {Arnouts}, St{\'e}phane and {Gwyn}, Stephen and {Marchesini}, Danilo and {Sajina}, Anna},
        title = "{The Growth of Galaxy Stellar Haloes over 0.2 {\ensuremath{\leq}} z {\ensuremath{\leq}} 1.1}",
      journal = {\apj},
         year = 2025,
        month = aug,
       volume = {989},
       number = {1},
          eid = {107},
        pages = {107},
          doi = {10.3847/1538-4357/ade9a8},
archivePrefix = {arXiv},
       eprint = {2412.03662},
 primaryClass = {astro-ph.GA},
       adsurl = {https://ui.adsabs.harvard.edu/abs/2025ApJ...989..107W}
}

@ARTICLE{Geha_2017,
       author = {{Geha}, Marla and {Wechsler}, Risa H. and {Mao}, Yao-Yuan and {Tollerud}, Erik J. and {Weiner}, Benjamin and {Bernstein}, Rebecca and {Hoyle}, Ben and {Marchi}, Sebastian and {Marshall}, Phil J. and {Mu{\~n}oz}, Ricardo and {Lu}, Yu},
        title = "{The SAGA Survey. I. Satellite Galaxy Populations around Eight Milky Way Analogs}",
      journal = {\apj},
         year = 2017,
        month = sep,
       volume = {847},
       number = {1},
          eid = {4},
        pages = {4},
          doi = {10.3847/1538-4357/aa8626},
archivePrefix = {arXiv},
       eprint = {1705.06743},
 primaryClass = {astro-ph.GA},
       adsurl = {https://ui.adsabs.harvard.edu/abs/2017ApJ...847....4G}
}

@ARTICLE{Slater_2014,
       author = {{Slater}, Colin T. and {Bell}, Eric F.},
        title = "{The Mass Dependence of Dwarf Satellite Galaxy Quenching}",
      journal = {\apj},
         year = 2014,
        month = sep,
       volume = {792},
       number = {2},
          eid = {141},
        pages = {141},
          doi = {10.1088/0004-637X/792/2/141},
archivePrefix = {arXiv},
       eprint = {1407.6006},
 primaryClass = {astro-ph.GA},
       adsurl = {https://ui.adsabs.harvard.edu/abs/2014ApJ...792..141S}
}

@ARTICLE{Scannapieco_2008,
       author = {{Scannapieco}, Cecilia and {Tissera}, Patricia B. and {White}, Simon D.~M. and {Springel}, Volker},
        title = "{Effects of supernova feedback on the formation of galaxy discs}",
      journal = {\mnras},
         year = 2008,
        month = sep,
       volume = {389},
       number = {3},
        pages = {1137-1149},
          doi = {10.1111/j.1365-2966.2008.13678.x},
archivePrefix = {arXiv},
       eprint = {0804.3795},
 primaryClass = {astro-ph},
       adsurl = {https://ui.adsabs.harvard.edu/abs/2008MNRAS.389.1137S}
}

@BOOK{Matteucci_2012,
       author = {{Matteucci}, Francesca},
        title = "{Chemical Evolution of Galaxies}",
         year = 2012,
          doi = {10.1007/978-3-642-22491-1},
       adsurl = {https://ui.adsabs.harvard.edu/abs/2012ceg..book.....M}
}

@ARTICLE{Font_2020,
       author = {{Font}, Andreea S. and {McCarthy}, Ian G. and {Poole-Mckenzie}, Robert and {Stafford}, Sam G. and {Brown}, Shaun T. and {Schaye}, Joop and {Crain}, Robert A. and {Theuns}, Tom and {Schaller}, Matthieu},
        title = "{The ARTEMIS simulations: stellar haloes of Milky Way-mass galaxies}",
      journal = {\mnras},
         year = 2020,
        month = oct,
       volume = {498},
       number = {2},
        pages = {1765-1785},
          doi = {10.1093/mnras/staa2463},
archivePrefix = {arXiv},
       eprint = {2004.01914},
 primaryClass = {astro-ph.GA},
       adsurl = {https://ui.adsabs.harvard.edu/abs/2020MNRAS.498.1765F}
}

@ARTICLE{Planck2014,
       author = {{Planck Collaboration} and {Ade}, P.~A.~R. and {Aghanim}, N. and {Armitage-Caplan}, C. and {Arnaud}, M. and {Ashdown}, M. and {Atrio-Barandela}, F. and {Aumont}, J. and {Baccigalupi}, C. and {Banday}, A.~J. and et al.},
        title = "{Planck 2013 results. XVI. Cosmological parameters}",
      journal = {\aap},
         year = 2014,
        month = nov,
       volume = {571},
          eid = {A16},
        pages = {A16},
          doi = {10.1051/0004-6361/201321591},
archivePrefix = {arXiv},
       eprint = {1303.5076},
 primaryClass = {astro-ph.CO},
       adsurl = {https://ui.adsabs.harvard.edu/abs/2014A\&A...571A..16P}
}

@ARTICLE{Freeman_2002,
       author = {{Freeman}, Ken and {Bland-Hawthorn}, Joss},
        title = "{The New Galaxy: Signatures of Its Formation}",
      journal = {\araa},
         year = 2002,
        month = jan,
       volume = {40},
        pages = {487-537},
          doi = {10.1146/annurev.astro.40.060401.093840},
archivePrefix = {arXiv},
       eprint = {astro-ph/0208106},
 primaryClass = {astro-ph},
       adsurl = {https://ui.adsabs.harvard.edu/abs/2002ARA&A..40..487F}
}

@ARTICLE{Proctor_2024,
       author = {{Proctor}, Katy L. and {Ludlow}, Aaron D. and {Lagos}, Claudia del P. and {Robotham}, Aaron S.~G.},
        title = "{The weak connection between the stellar haloes and merger histories of Milky Way-mass galaxies}",
      journal = {\mnras},
         year = 2025,
        month = sep,
       volume = {542},
       number = {2},
        pages = {1673-1683},
          doi = {10.1093/mnras/staf1339},
archivePrefix = {arXiv},
       eprint = {2407.11444},
 primaryClass = {astro-ph.GA},
       adsurl = {https://ui.adsabs.harvard.edu/abs/2025MNRAS.542.1673P}
}

@ARTICLE{Canas_2020,
       author = {{Ca{\~n}as}, Rodrigo and {Lagos}, Claudia del P. and {Elahi}, Pascal J. and {Power}, Chris and {Welker}, Charlotte and {Dubois}, Yohan and {Pichon}, Christophe},
        title = "{From stellar haloes to intracluster light: the physics of the Intra-Halo Stellar Component in cosmological hydrodynamical simulations}",
      journal = {\mnras},
         year = 2020,
        month = may,
       volume = {494},
       number = {3},
        pages = {4314-4333},
          doi = {10.1093/mnras/staa1027},
archivePrefix = {arXiv},
       eprint = {1908.02945},
 primaryClass = {astro-ph.GA},
       adsurl = {https://ui.adsabs.harvard.edu/abs/2020MNRAS.494.4314C}
}

@ARTICLE{Pulsoni_2020,
       author = {{Pulsoni}, C. and {Gerhard}, O. and {Arnaboldi}, M. and {Pillepich}, A. and {Nelson}, D. and {Hernquist}, L. and {Springel}, V.},
        title = "{The stellar halos of ETGs in the IllustrisTNG simulations: The photometric and kinematic diversity of galaxies at large radii}",
      journal = {\aap},
         year = 2020,
        month = sep,
       volume = {641},
          eid = {A60},
        pages = {A60},
          doi = {10.1051/0004-6361/202038253},
archivePrefix = {arXiv},
       eprint = {2004.13042},
 primaryClass = {astro-ph.GA},
       adsurl = {https://ui.adsabs.harvard.edu/abs/2020A&A...641A..60P}
}

@ARTICLE{Yun_Pu_2025,
       author = {{Pu}, Sy-Yun and {Cooper}, Andrew P. and {Grand}, Robert J.~J. and {G{\'o}mez}, Facundo A. and {Monachesi}, Antonela},
        title = "{Progenitor Diversity in the Accreted Stellar Halos of Milky Way{\textendash}like Galaxies}",
      journal = {\apj},
         year = 2025,
        month = feb,
       volume = {980},
       number = {1},
          eid = {63},
        pages = {63},
          doi = {10.3847/1538-4357/ada382},
archivePrefix = {arXiv},
       eprint = {2410.13491},
 primaryClass = {astro-ph.GA},
       adsurl = {https://ui.adsabs.harvard.edu/abs/2025ApJ...980...63P}
}

@ARTICLE{Cooper_2025,
       author = {{Cooper}, Andrew P. and {Frenk}, Carlos S. and {Hellwing}, Wojciech A. and {Bose}, Sownak},
        title = "{Simulations of the accreted stellar haloes of low-mass field galaxies}",
      journal = {\mnras},
         year = 2025,
        month = jul,
       volume = {540},
       number = {3},
        pages = {2049-2080},
          doi = {10.1093/mnras/staf833},
archivePrefix = {arXiv},
       eprint = {2501.13317},
 primaryClass = {astro-ph.GA},
       adsurl = {https://ui.adsabs.harvard.edu/abs/2025MNRAS.540.2049C}
}

@ARTICLE{Font_2006,
       author = {{Font}, Andreea S. and {Johnston}, Kathryn V. and {Bullock}, James S. and {Robertson}, Brant E.},
        title = "{Phase-Space Distributions of Chemical Abundances in Milky Way-Type Galaxy Halos}",
      journal = {\apj},
         year = 2006,
        month = aug,
       volume = {646},
       number = {2},
        pages = {886-898},
          doi = {10.1086/505131},
archivePrefix = {arXiv},
       eprint = {astro-ph/0512611},
 primaryClass = {astro-ph},
       adsurl = {https://ui.adsabs.harvard.edu/abs/2006ApJ...646..886F}
}

@ARTICLE{Robertson_2005,
       author = {{Robertson}, Brant and {Bullock}, James S. and {Font}, Andreea S. and {Johnston}, Kathryn V. and {Hernquist}, Lars},
        title = "{{\ensuremath{\Lambda}} Cold Dark Matter, Stellar Feedback, and the Galactic Halo Abundance Pattern}",
      journal = {\apj},
         year = 2005,
        month = oct,
       volume = {632},
       number = {2},
        pages = {872-881},
          doi = {10.1086/452619},
archivePrefix = {arXiv},
       eprint = {astro-ph/0501398},
 primaryClass = {astro-ph},
       adsurl = {https://ui.adsabs.harvard.edu/abs/2005ApJ...632..872R}
}

@ARTICLE{Smercina_2022,
       author = {{Smercina}, Adam and {Bell}, Eric F. and {Samuel}, Jenna and {D'Souza}, Richard},
        title = "{Relating the Diverse Merger Histories and Satellite Populations of Nearby Galaxies}",
      journal = {\apj},
         year = 2022,
        month = may,
       volume = {930},
       number = {1},
          eid = {69},
        pages = {69},
          doi = {10.3847/1538-4357/ac5d56},
archivePrefix = {arXiv},
       eprint = {2107.04591},
 primaryClass = {astro-ph.GA},
       adsurl = {https://ui.adsabs.harvard.edu/abs/2022ApJ...930...69S}
}

@ARTICLE{Bullock_2005,
       author = {{Bullock}, James S. and {Johnston}, Kathryn V.},
        title = "{Tracing Galaxy Formation with Stellar Halos. I. Methods}",
      journal = {\apj},
         year = 2005,
        month = dec,
       volume = {635},
       number = {2},
        pages = {931-949},
          doi = {10.1086/497422},
archivePrefix = {arXiv},
       eprint = {astro-ph/0506467},
 primaryClass = {astro-ph},
       adsurl = {https://ui.adsabs.harvard.edu/abs/2005ApJ...635..931B}
}

@ARTICLE{Zolotov_2009,
       author = {{Zolotov}, Adi and {Willman}, Beth and {Brooks}, Alyson M. and {Governato}, Fabio and {Brook}, Chris B. and {Hogg}, David W. and {Quinn}, Tom and {Stinson}, Greg},
        title = "{The Dual Origin of Stellar Halos}",
      journal = {\apj},
         year = 2009,
        month = sep,
       volume = {702},
       number = {2},
        pages = {1058-1067},
          doi = {10.1088/0004-637X/702/2/1058},
archivePrefix = {arXiv},
       eprint = {0904.3333},
 primaryClass = {astro-ph.GA},
       adsurl = {https://ui.adsabs.harvard.edu/abs/2009ApJ...702.1058Z}
}

@ARTICLE{Tissera_2012,
       author = {{Tissera}, Patricia B. and {White}, Simon D.M. and {Scannapieco}, Cecilia},
        title = "{Chemical signatures of formation processes in the stellar populations of simulated galaxies}",
      journal = {\mnras},
         year = 2012,
        month = feb,
       volume = {420},
       number = {1},
        pages = {255-270},
          doi = {10.1111/j.1365-2966.2011.20028.x},
archivePrefix = {arXiv},
       eprint = {1110.5864},
 primaryClass = {astro-ph.CO},
       adsurl = {https://ui.adsabs.harvard.edu/abs/2012MNRAS.420..255T}
}

@ARTICLE{Wittig_2025,
       author = {{Wittig}, Ole and {Ramesh}, Rahul and {Nelson}, Dylan},
        title = "{Tracing the cosmological origin of gas that fuels in situ star formation in TNG50 galaxies}",
      journal = {\aap},
         year = 2025,
        month = mar,
       volume = {695},
          eid = {A121},
        pages = {A121},
          doi = {10.1051/0004-6361/202453101},
archivePrefix = {arXiv},
       eprint = {2503.00111},
 primaryClass = {astro-ph.GA},
       adsurl = {https://ui.adsabs.harvard.edu/abs/2025A&A...695A.121W}
}

@ARTICLE{Harmsen_2023,
       author = {{Harmsen}, Benjamin and {Bell}, Eric F. and {D'Souza}, Richard and {Monachesi}, Antonela and {de Jong}, Roelof S. and {Smercina}, Adam and {Jang}, In Sung and {Holwerda}, Benne W.},
        title = "{Constraining the assembly time of the stellar haloes of nearby Milky Way-mass galaxies through AGB populations}",
      journal = {\mnras},
         year = 2023,
        month = nov,
       volume = {525},
       number = {3},
        pages = {4497-4514},
          doi = {10.1093/mnras/stad2480},
archivePrefix = {arXiv},
       eprint = {2308.11499},
 primaryClass = {astro-ph.GA},
       adsurl = {https://ui.adsabs.harvard.edu/abs/2023MNRAS.525.4497H}
}

@ARTICLE{Celiz_2025,
       author = {{Celiz}, Bruno M. and {Navarro}, Julio F. and {Abadi}, Mario G.},
        title = "{Accreted stars and stellar haloes of simulated galaxies in TNG50}",
      journal = {\aap},
         year = 2025,
        month = nov,
       volume = {704},
          eid = {A57},
        pages = {A57},
          doi = {10.1051/0004-6361/202556633},
archivePrefix = {arXiv},
       eprint = {2510.18971},
 primaryClass = {astro-ph.GA},
       adsurl = {https://ui.adsabs.harvard.edu/abs/2025A&A...704A..57C}
}

@ARTICLE{Vera-Casanova_2025,
       author = {{Vera-Casanova}, Alex and {Monsalves Gonzalez}, Nicolas and {G{\'o}mez}, Facundo A. and {Jaque Arancibia}, Marcelo and {Fontirroig}, Valentina and {Pallero}, Diego and {Pakmor}, R{\"u}diger and {van de Voort}, Freeke and {Grand}, Robert J.~J. and {Bieri}, Rebekka and {Marinacci}, Federico},
        title = "{Stream automatic detection with convolutional neural networks}",
      journal = {\aap},
         year = 2025,
        month = dec,
       volume = {704},
          eid = {A53},
        pages = {A53},
          doi = {10.1051/0004-6361/202554688},
archivePrefix = {arXiv},
       eprint = {2503.17202},
 primaryClass = {astro-ph.GA},
       adsurl = {https://ui.adsabs.harvard.edu/abs/2025A&A...704A..53V}
}

@ARTICLE{Fattahi_2020,
       author = {{Fattahi}, Azadeh and {Deason}, Alis J. and {Frenk}, Carlos S. and {Simpson}, Christine M. and {G{\'o}mez}, Facundo A. and {Grand}, Robert J.~J. and {Monachesi}, Antonela and {Marinacci}, Federico and {Pakmor}, R{\"u}diger},
        title = "{A tale of two populations: surviving and destroyed dwarf galaxies and the build-up of the Milky Way's stellar halo}",
      journal = {\mnras},
         year = 2020,
        month = oct,
       volume = {497},
       number = {4},
        pages = {4459-4471},
          doi = {10.1093/mnras/staa2221},
archivePrefix = {arXiv},
       eprint = {2002.12043},
 primaryClass = {astro-ph.GA},
       adsurl = {https://ui.adsabs.harvard.edu/abs/2020MNRAS.497.4459F}
}

@ARTICLE{Monachesi_2016b,
       author = {{Monachesi}, Antonela and {G{\'o}mez}, Facundo A. and {Grand}, Robert J.~J. and {Kauffmann}, Guinevere and {Marinacci}, Federico and {Pakmor}, R{\"u}diger and {Springel}, Volker and {Frenk}, Carlos S.},
        title = "{On the stellar halo metallicity profile of Milky Way-like galaxies in the Auriga simulations}",
      journal = {\mnras},
         year = 2016,
        month = jun,
       volume = {459},
       number = {1},
        pages = {L46-L50},
          doi = {10.1093/mnrasl/slw052},
archivePrefix = {arXiv},
       eprint = {1512.03064},
 primaryClass = {astro-ph.GA},
       adsurl = {https://ui.adsabs.harvard.edu/abs/2016MNRAS.459L..46M}
}

@ARTICLE{Jang_2020,
       author = {{Jang}, In Sung and {de Jong}, Roelof S. and {Holwerda}, Benne W. and {Monachesi}, Antonela and {Bell}, Eric F. and {Bailin}, Jeremy},
        title = "{Tracing the anemic stellar halo of M 101}",
      journal = {\aap},
         year = 2020,
        month = may,
       volume = {637},
          eid = {A8},
        pages = {A8},
          doi = {10.1051/0004-6361/201936994},
archivePrefix = {arXiv},
       eprint = {2001.12007},
 primaryClass = {astro-ph.GA},
       adsurl = {https://ui.adsabs.harvard.edu/abs/2020A\&A...637A...8J}
}

@ARTICLE{Elias_2018,
       author = {{Elias}, Lydia M. and {Sales}, Laura V. and {Creasey}, Peter and {Cooper}, Michael C. and {Bullock}, James S. and {Rich}, R. Michael and {Hernquist}, Lars},
        title = "{Stellar halos in Illustris: probing the histories of Milky Way-mass galaxies}",
      journal = {\mnras},
         year = 2018,
        month = sep,
       volume = {479},
       number = {3},
        pages = {4004-4016},
          doi = {10.1093/mnras/sty1718},
archivePrefix = {arXiv},
       eprint = {1801.07273},
 primaryClass = {astro-ph.GA},
       adsurl = {https://ui.adsabs.harvard.edu/abs/2018MNRAS.479.4004E}
}

@ARTICLE{Amorisco_2017,
       author = {{Amorisco}, Nicola C.},
        title = "{The accreted stellar halo as a window on halo assembly in L$^{*}$ galaxies}",
      journal = {\mnras},
         year = 2017,
        month = jul,
       volume = {469},
       number = {1},
        pages = {L48-L52},
          doi = {10.1093/mnrasl/slx044},
archivePrefix = {arXiv},
       eprint = {1701.02741},
 primaryClass = {astro-ph.GA},
       adsurl = {https://ui.adsabs.harvard.edu/abs/2017MNRAS.469L..48A}
}

@ARTICLE{Molla_2019,
       author = {{Moll{\'a}}, M. and {D{\'\i}az}, {\'A}. I. and {Cavichia}, O. and {Gibson}, B.~K. and {Maciel}, W.~J. and {Costa}, R.~D.~D. and {Ascasibar}, Y. and {Few}, C.~G.},
        title = "{The time evolution of the Milky Way's oxygen abundance gradient}",
      journal = {\mnras},
         year = 2019,
        month = jan,
       volume = {482},
       number = {3},
        pages = {3071-3088},
          doi = {10.1093/mnras/sty2877},
archivePrefix = {arXiv},
       eprint = {1810.09182},
 primaryClass = {astro-ph.GA},
       adsurl = {https://ui.adsabs.harvard.edu/abs/2019MNRAS.482.3071M}
}

@ARTICLE{Survey_SDSS_2000,
       author = {{York}, Donald G. and {Adelman}, J. and {Anderson}, John E., Jr. and {Anderson}, Scott F. and {Annis}, James and {Bahcall}, Neta A. and {Bakken}, J.A. and {Barkhouser}, Robert and {Bastian}, Steven and {Berman}, Eileen and {Boroski}, William N. and {Bracker}, Steve and {Briegel}, Charlie and {Briggs}, John W. and {Brinkmann}, J. and {Brunner}, Robert and {Burles}, Scott and {Carey}, Larry and {Carr}, Michael A. and {Castander}, Francisco J. and {Chen}, Bing and {Colestock}, Patrick L. and {Connolly}, A.J. and {Crocker}, J.H. and {Csabai}, Istv{\'a}n and {Czarapata}, Paul C. and {Davis}, John Eric and {Doi}, Mamoru and {Dombeck}, Tom and {Eisenstein}, Daniel and {Ellman}, Nancy and {Elms}, Brian R. and {Evans}, Michael L. and {Fan}, Xiaohui and {Federwitz}, Glenn R. and {Fiscelli}, Larry and {Friedman}, Scott and {Frieman}, Joshua A. and {Fukugita}, Masataka and {Gillespie}, Bruce and {Gunn}, James E. and {Gurbani}, Vijay K. and {de Haas}, Ernst and {Haldeman}, Merle and {Harris}, Frederick H. and {Hayes}, J. and {Heckman}, Timothy M. and {Hennessy}, G.S. and {Hindsley}, Robert B. and {Holm}, Scott and {Holmgren}, Donald J. and {Huang}, Chi-hao and {Hull}, Charles and {Husby}, Don and {Ichikawa}, Shin-Ichi and {Ichikawa}, Takashi and {Ivezi{\'c}}, {\v{Z}}eljko and {Kent}, Stephen and {Kim}, Rita S.J. and {Kinney}, E. and {Klaene}, Mark and {Kleinman}, A.N. and {Kleinman}, S. and {Knapp}, G.R. and {Korienek}, John and {Kron}, Richard G. and {Kunszt}, Peter Z. and {Lamb}, D.Q. and {Lee}, B. and {Leger}, R. French and {Limmongkol}, Siriluk and {Lindenmeyer}, Carl and {Long}, Daniel C. and {Loomis}, Craig and {Loveday}, Jon and {Lucinio}, Rich and {Lupton}, Robert H. and {MacKinnon}, Bryan and {Mannery}, Edward J. and {Mantsch}, P.M. and {Margon}, Bruce and {McGehee}, Peregrine and {McKay}, Timothy A. and {Meiksin}, Avery and {Merelli}, Aronne and {Monet}, David G. and {Munn}, Jeffrey A. and {Narayanan}, Vijay K. and {Nash}, Thomas and {Neilsen}, Eric and {Neswold}, Rich and {Newberg}, Heidi Jo and {Nichol}, R.C. and {Nicinski}, Tom and {Nonino}, Mario and {Okada}, Norio and {Okamura}, Sadanori and {Ostriker}, Jeremiah P. and {Owen}, Russell and {Pauls}, A. George and {Peoples}, John and {Peterson}, R.L. and {Petravick}, Donald and {Pier}, Jeffrey R. and {Pope}, Adrian and {Pordes}, Ruth and {Prosapio}, Angela and {Rechenmacher}, Ron and {Quinn}, Thomas R. and {Richards}, Gordon T. and {Richmond}, Michael W. and {Rivetta}, Claudio H. and {Rockosi}, Constance M. and {Ruthmansdorfer}, Kurt and {Sandford}, Dale and {Schlegel}, David J. and {Schneider}, Donald P. and {Sekiguchi}, Maki and {Sergey}, Gary and {Shimasaku}, Kazuhiro and {Siegmund}, Walter A. and {Smee}, Stephen and {Smith}, J. Allyn and {Snedden}, S. and {Stone}, R. and {Stoughton}, Chris and {Strauss}, Michael A. and {Stubbs}, Christopher and {SubbaRao}, Mark and {Szalay}, Alexander S. and {Szapudi}, Istvan and {Szokoly}, Gyula P. and {Thakar}, Anirudda R. and {Tremonti}, Christy and {Tucker}, Douglas L. and {Uomoto}, Alan and {Vanden Berk}, Dan and {Vogeley}, Michael S. and {Waddell}, Patrick and {Wang}, Shu-i. and {Watanabe}, Masaru and {Weinberg}, David H. and {Yanny}, Brian and {Yasuda}, Naoki and {SDSS Collaboration}},
        title = "{The Sloan Digital Sky Survey: Technical Summary}",
      journal = {\aj},
         year = 2000,
        month = sep,
       volume = {120},
       number = {3},
        pages = {1579-1587},
          doi = {10.1086/301513},
archivePrefix = {arXiv},
       eprint = {astro-ph/0006396},
 primaryClass = {astro-ph},
       adsurl = {https://ui.adsabs.harvard.edu/abs/2000AJ....120.1579Y}
}

@ARTICLE{Smercina_2020,
       author = {{Smercina}, Adam and {Bell}, Eric F. and {Price}, Paul A. and {Slater}, Colin T. and {D'Souza}, Richard and {Bailin}, Jeremy and {de Jong}, Roelof S. and {Jang}, In Sung and {Monachesi}, Antonela and {Nidever}, David},
        title = "{The Saga of M81: Global View of a Massive Stellar Halo in Formation}",
      journal = {\apj},
         year = 2020,
        month = dec,
       volume = {905},
       number = {1},
          eid = {60},
        pages = {60},
          doi = {10.3847/1538-4357/abc485},
archivePrefix = {arXiv},
       eprint = {1910.14672},
 primaryClass = {astro-ph.GA},
       adsurl = {https://ui.adsabs.harvard.edu/abs/2020ApJ...905...60S}
}

@ARTICLE{Di_Matteo_2019,
       author = {{Di Matteo}, P. and {Haywood}, M. and {Lehnert}, M.~D. and {Katz}, D. and {Khoperskov}, S. and {Snaith}, O.~N. and {G{\'o}mez}, A. and {Robichon}, N.},
        title = "{The Milky Way has no in-situ halo other than the heated thick disc. Composition of the stellar halo and age-dating the last significant merger with Gaia DR2 and APOGEE}",
      journal = {\aap},
         year = 2019,
        month = dec,
       volume = {632},
          eid = {A4},
        pages = {A4},
          doi = {10.1051/0004-6361/201834929},
archivePrefix = {arXiv},
       eprint = {1812.08232},
 primaryClass = {astro-ph.GA},
       adsurl = {https://ui.adsabs.harvard.edu/abs/2019A&A...632A...4D}
}

@ARTICLE{Giribaldi_2023,
       author = {{Giribaldi}, R.~E. and {Smiljanic}, R.},
        title = "{Chronology of the chemical enrichment of the old Galactic stellar populations}",
      journal = {\aap},
         year = 2023,
        month = may,
       volume = {673},
          eid = {A18},
        pages = {A18},
          doi = {10.1051/0004-6361/202245404},
archivePrefix = {arXiv},
       eprint = {2302.09640},
 primaryClass = {astro-ph.GA},
       adsurl = {https://ui.adsabs.harvard.edu/abs/2023A&A...673A..18G}
}

@ARTICLE{Gallart_2019,
       author = {{Gallart}, Carme and {Bernard}, Edouard J. and {Brook}, Chris B. and {Ruiz-Lara}, Tom{\'a}s and {Cassisi}, Santi and {Hill}, Vanessa and {Monelli}, Matteo},
        title = "{Uncovering the birth of the Milky Way through accurate stellar ages with Gaia}",
      journal = {Nature Astronomy},
         year = 2019,
        month = jul,
       volume = {3},
        pages = {932-939},
          doi = {10.1038/s41550-019-0829-5},
archivePrefix = {arXiv},
       eprint = {1901.02900},
 primaryClass = {astro-ph.GA},
       adsurl = {https://ui.adsabs.harvard.edu/abs/2019NatAs...3..932G}
}

@ARTICLE{Belokurov_2018,
       author = {{Belokurov}, V. and {Erkal}, D. and {Evans}, N.W. and {Koposov}, S.E. and {Deason}, A.J.},
        title = "{Co-formation of the disc and the stellar halo}",
      journal = {\mnras},
         year = 2018,
        month = jul,
       volume = {478},
       number = {1},
        pages = {611-619},
          doi = {10.1093/mnras/sty982},
archivePrefix = {arXiv},
       eprint = {1802.03414},
 primaryClass = {astro-ph.GA},
       adsurl = {https://ui.adsabs.harvard.edu/abs/2018MNRAS.478..611B}
}

@ARTICLE{Helmi_2018,
       author = {{Helmi}, Amina and {Babusiaux}, Carine and {Koppelman}, Helmer H. and {Massari}, Davide and {Veljanoski}, Jovan and {Brown}, Anthony G.A.},
        title = "{The merger that led to the formation of the Milky Way's inner stellar halo and thick disk}",
      journal = {\nat},
         year = 2018,
        month = oct,
       volume = {563},
       number = {7729},
        pages = {85-88},
          doi = {10.1038/s41586-018-0625-x},
archivePrefix = {arXiv},
       eprint = {1806.06038},
 primaryClass = {astro-ph.GA},
       adsurl = {https://ui.adsabs.harvard.edu/abs/2018Natur.563...85H}
}

@ARTICLE{Tissera_2013,
       author = {{Tissera}, Patricia B. and {Scannapieco}, Cecilia and {Beers}, Timothy C. and {Carollo}, Daniela},
        title = "{Stellar haloes of simulated Milky-Way-like galaxies: chemical and kinematic properties}",
      journal = {\mnras},
         year = 2013,
        month = jul,
       volume = {432},
       number = {4},
        pages = {3391-3400},
          doi = {10.1093/mnras/stt691},
archivePrefix = {arXiv},
       eprint = {1301.1301},
 primaryClass = {astro-ph.GA},
       adsurl = {https://ui.adsabs.harvard.edu/abs/2013MNRAS.432.3391T}
}

@ARTICLE{Tissera_2014,
       author = {{Tissera}, Patricia B. and {Beers}, Timothy C. and {Carollo}, Daniela and {Scannapieco}, Cecilia},
        title = "{Stellar haloes in Milky Way mass galaxies: from the inner to the outer haloes}",
      journal = {\mnras},
         year = 2014,
        month = apr,
       volume = {439},
       number = {3},
        pages = {3128-3138},
          doi = {10.1093/mnras/stu181},
archivePrefix = {arXiv},
       eprint = {1309.3609},
 primaryClass = {astro-ph.CO},
       adsurl = {https://ui.adsabs.harvard.edu/abs/2014MNRAS.439.3128T}
}

@ARTICLE{Rey_2021,
       author = {{Rey}, Martin P. and {Starkenburg}, Tjitske K.},
        title = "{How cosmological merger histories shape the diversity of stellar haloes}",
      journal = {\mnras},
         year = 2022,
        month = mar,
       volume = {510},
       number = {3},
        pages = {4208-4224},
          doi = {10.1093/mnras/stab3709},
archivePrefix = {arXiv},
       eprint = {2106.09729},
 primaryClass = {astro-ph.GA},
       adsurl = {https://ui.adsabs.harvard.edu/abs/2022MNRAS.510.4208R}
}

@ARTICLE{Gomez_2010,
       author = {{G{\'o}mez}, Facundo A. and {Helmi}, Amina},
        title = "{On the identification of substructure in phase space using orbital frequencies}",
      journal = {\mnras},
         year = 2010,
        month = feb,
       volume = {401},
       number = {4},
        pages = {2285-2298},
          doi = {10.1111/j.1365-2966.2009.15841.x},
archivePrefix = {arXiv},
       eprint = {0904.1377},
 primaryClass = {astro-ph.GA},
       adsurl = {https://ui.adsabs.harvard.edu/abs/2010MNRAS.401.2285G}
}

@ARTICLE{Harmsen_2017,
       author = {{Harmsen}, Benjamin and {Monachesi}, Antonela and {Bell}, Eric F. and {de Jong}, Roelof S. and {Bailin}, Jeremy and {Radburn-Smith}, David J. and {Holwerda}, Benne W.},
        title = "{Diverse stellar haloes in nearby Milky Way mass disc galaxies}",
      journal = {\mnras},
         year = 2017,
        month = apr,
       volume = {466},
       number = {2},
        pages = {1491-1512},
          doi = {10.1093/mnras/stw2992},
archivePrefix = {arXiv},
       eprint = {1611.05448},
 primaryClass = {astro-ph.GA},
       adsurl = {https://ui.adsabs.harvard.edu/abs/2017MNRAS.466.1491H}
}

@ARTICLE{DSouza_2018,
       author = {{D'Souza}, Richard and {Bell}, Eric F.},
        title = "{The masses and metallicities of stellar haloes reflect galactic merger histories}",
      journal = {\mnras},
         year = 2018,
        month = mar,
       volume = {474},
       number = {4},
        pages = {5300-5318},
          doi = {10.1093/mnras/stx3081},
archivePrefix = {arXiv},
       eprint = {1705.08442},
 primaryClass = {astro-ph.GA},
       adsurl = {https://ui.adsabs.harvard.edu/abs/2018MNRAS.474.5300D}
}

@ARTICLE{Genel_2014_Illustris,
       author = {{Genel}, Shy and {Vogelsberger}, Mark and {Springel}, Volker and {Sijacki}, Debora and {Nelson}, Dylan and {Snyder}, Greg and {Rodriguez-Gomez}, Vicente and {Torrey}, Paul and {Hernquist}, Lars},
        title = "{Introducing the Illustris project: the evolution of galaxy populations across cosmic time}",
      journal = {\mnras},
         year = 2014,
        month = nov,
       volume = {445},
       number = {1},
        pages = {175-200},
          doi = {10.1093/mnras/stu1654},
archivePrefix = {arXiv},
       eprint = {1405.3749},
 primaryClass = {astro-ph.CO},
       adsurl = {https://ui.adsabs.harvard.edu/abs/2014MNRAS.445..175G}
}

@ARTICLE{Vogelsberger_2013_Illustris,
       author = {{Torrey}, Paul and {Vogelsberger}, Mark and {Genel}, Shy and {Sijacki}, Debora and {Springel}, Volker and {Hernquist}, Lars},
        title = "{A model for cosmological simulations of galaxy formation physics: multi-epoch validation}",
      journal = {\mnras},
         year = 2014,
        month = mar,
       volume = {438},
       number = {3},
        pages = {1985-2004},
          doi = {10.1093/mnras/stt2295},
archivePrefix = {arXiv},
       eprint = {1305.4931},
 primaryClass = {astro-ph.CO},
       adsurl = {https://ui.adsabs.harvard.edu/abs/2014MNRAS.438.1985T}
}

@ARTICLE{Khoperskov_2023c,
       author = {{Khoperskov}, Sergey and {Minchev}, Ivan and {Libeskind}, Noam and {Belokurov}, Vasily and {Steinmetz}, Matthias and {Gomez}, Facundo A. and {Grand}, Robert J.~J. and {Hoffman}, Yehuda and {Knebe}, Alexander and {Sorce}, Jenny G. and {Spaare}, Martin and {Tempel}, Elmo and {Vogelsberger}, Mark},
        title = "{The stellar halo in Local Group Hestia simulations. III. Chemical abundance relations for accreted and in situ stars}",
      journal = {\aap},
         year = 2023,
        month = sep,
       volume = {677},
          eid = {A91},
        pages = {A91},
          doi = {10.1051/0004-6361/202244234},
archivePrefix = {arXiv},
       eprint = {2206.05491},
 primaryClass = {astro-ph.GA},
       adsurl = {https://ui.adsabs.harvard.edu/abs/2023A\&A...677A..91K}
}

@ARTICLE{Deason_2016,
       author = {{Deason}, Alis J. and {Mao}, Yao-Yuan and {Wechsler}, Risa H.},
        title = "{The Eating Habits of Milky Way-mass Halos: Destroyed Dwarf Satellites and the Metallicity Distribution of Accreted Stars}",
      journal = {\apj},
         year = 2016,
        month = apr,
       volume = {821},
       number = {1},
          eid = {5},
        pages = {5},
          doi = {10.3847/0004-637X/821/1/5},
archivePrefix = {arXiv},
       eprint = {1601.07905},
 primaryClass = {astro-ph.GA},
       adsurl = {https://ui.adsabs.harvard.edu/abs/2016ApJ...821....5D}
}

@ARTICLE{Monachesi_2019,
       author = {{Monachesi}, Antonela and {G{\'o}mez}, Facundo A. and {Grand}, Robert J.~J. and {Simpson}, Christine M. and {Kauffmann}, Guinevere and {Bustamante}, Sebasti{\'a}n and {Marinacci}, Federico and {Pakmor}, R{\"u}diger and {Springel}, Volker and {Frenk}, Carlos S. and {White}, Simon D.~M. and {Tissera}, Patricia B.},
        title = "{The Auriga stellar haloes: connecting stellar population properties with accretion and merging history}",
      journal = {\mnras},
         year = 2019,
        month = may,
       volume = {485},
       number = {2},
        pages = {2589-2616},
          doi = {10.1093/mnras/stz538},
archivePrefix = {arXiv},
       eprint = {1804.07798},
 primaryClass = {astro-ph.GA},
       adsurl = {https://ui.adsabs.harvard.edu/abs/2019MNRAS.485.2589M}
}

@ARTICLE{Bell_2017,
       author = {{Bell}, Eric F. and {Monachesi}, Antonela and {Harmsen}, Benjamin and {de Jong}, Roelof S. and {Bailin}, Jeremy and {Radburn-Smith}, David J. and {D'Souza}, Richard and {Holwerda}, Benne W.},
        title = "{Galaxies Grow Their Bulges and Black Holes in Diverse Ways}",
      journal = {\apjl},
         year = 2017,
        month = mar,
       volume = {837},
       number = {1},
          eid = {L8},
        pages = {L8},
          doi = {10.3847/2041-8213/aa6158},
archivePrefix = {arXiv},
       eprint = {1702.06116},
 primaryClass = {astro-ph.GA},
       adsurl = {https://ui.adsabs.harvard.edu/abs/2017ApJ...837L...8B}
}

@ARTICLE{Tinsley_79,
       author = {{Tinsley}, B.~M.},
        title = "{Stellar lifetimes and abundance ratios in chemical evolution.}",
      journal = {\apj},
         year = 1979,
        month = may,
       volume = {229},
        pages = {1046-1056},
          doi = {10.1086/157039},
       adsurl = {https://ui.adsabs.harvard.edu/abs/1979ApJ...229.1046T}
}

@ARTICLE{Cooper_2015,
       author = {{Cooper}, Andrew P. and {Parry}, Owen H. and {Lowing}, Ben and {Cole}, Shaun and {Frenk}, Carlos},
        title = "{Formation of in situ stellar haloes in Milky Way-mass galaxies}",
      journal = {\mnras},
         year = 2015,
        month = dec,
       volume = {454},
       number = {3},
        pages = {3185-3199},
          doi = {10.1093/mnras/stv2057},
archivePrefix = {arXiv},
       eprint = {1501.04630},
 primaryClass = {astro-ph.GA},
       adsurl = {https://ui.adsabs.harvard.edu/abs/2015MNRAS.454.3185C}
}

@ARTICLE{Purcell_2010,
       author = {{Purcell}, Chris W. and {Bullock}, James S. and {Kazantzidis}, Stelios},
        title = "{Heated disc stars in the stellar halo}",
      journal = {\mnras},
         year = 2010,
        month = jun,
       volume = {404},
       number = {4},
        pages = {1711-1718},
          doi = {10.1111/j.1365-2966.2010.16429.x},
archivePrefix = {arXiv},
       eprint = {0910.5481},
 primaryClass = {astro-ph.GA},
       adsurl = {https://ui.adsabs.harvard.edu/abs/2010MNRAS.404.1711P}
}

@ARTICLE{Lequeux_1979,
       author = {{Lequeux}, J. and {Peimbert}, M. and {Rayo}, J.~F. and {Serrano}, A. and {Torres-Peimbert}, S.},
        title = "{Chemical Composition and Evolution of Irregular and Blue Compact Galaxies}",
      journal = {\aap},
         year = 1979,
        month = dec,
       volume = {80},
        pages = {155},
       adsurl = {https://ui.adsabs.harvard.edu/abs/1979A\&A....80..155L}
}

@ARTICLE{Tremonti_2004,
       author = {{Tremonti}, Christy A. and {Heckman}, Timothy M. and {Kauffmann}, Guinevere and {Brinchmann}, Jarle and {Charlot}, St{\'e}phane and {White}, Simon D.~M. and {Seibert}, Mark and {Peng}, Eric W. and {Schlegel}, David J. and {Uomoto}, Alan and {Fukugita}, Masataka and {Brinkmann}, Jon},
        title = "{The Origin of the Mass-Metallicity Relation: Insights from 53,000 Star-forming Galaxies in the Sloan Digital Sky Survey}",
      journal = {\apj},
         year = 2004,
        month = oct,
       volume = {613},
       number = {2},
        pages = {898-913},
          doi = {10.1086/423264},
archivePrefix = {arXiv},
       eprint = {astro-ph/0405537},
 primaryClass = {astro-ph},
       adsurl = {https://ui.adsabs.harvard.edu/abs/2004ApJ...613..898T}
}

@ARTICLE{Lee_2006,
       author = {{Lee}, Henry and {Skillman}, Evan D. and {Cannon}, John M. and {Jackson}, Dale C. and {Gehrz}, Robert D. and {Polomski}, Elisha F. and {Woodward}, Charles E.},
        title = "{On Extending the Mass-Metallicity Relation of Galaxies by 2.5 Decades in Stellar Mass}",
      journal = {\apj},
         year = 2006,
        month = aug,
       volume = {647},
       number = {2},
        pages = {970-983},
          doi = {10.1086/505573},
archivePrefix = {arXiv},
       eprint = {astro-ph/0605036},
 primaryClass = {astro-ph},
       adsurl = {https://ui.adsabs.harvard.edu/abs/2006ApJ...647..970L}
}

@ARTICLE{Kirby_2013,
       author = {{Kirby}, Evan N. and {Cohen}, Judith G. and {Guhathakurta}, Puragra and {Cheng}, Lucy and {Bullock}, James S. and {Gallazzi}, Anna},
        title = "{The Universal Stellar Mass-Stellar Metallicity Relation for Dwarf Galaxies}",
      journal = {\apj},
         year = 2013,
        month = dec,
       volume = {779},
       number = {2},
          eid = {102},
        pages = {102},
          doi = {10.1088/0004-637X/779/2/102},
archivePrefix = {arXiv},
       eprint = {1310.0814},
 primaryClass = {astro-ph.GA},
       adsurl = {https://ui.adsabs.harvard.edu/abs/2013ApJ...779..102K}
}

@ARTICLE{Springel_2001,
       author = {{Springel}, Volker and {Yoshida}, Naoki and {White}, Simon D.~M.},
        title = "{GADGET: a code for collisionless and gasdynamical cosmological simulations}",
      journal = {\na},
         year = 2001,
        month = apr,
       volume = {6},
       number = {2},
        pages = {79-117},
          doi = {10.1016/S1384-1076(01)00042-2},
archivePrefix = {arXiv},
       eprint = {astro-ph/0003162},
 primaryClass = {astro-ph},
       adsurl = {https://ui.adsabs.harvard.edu/abs/2001NewA....6...79S}
}

@ARTICLE{Springel_2003,
       author = {{Springel}, Volker and {Hernquist}, Lars},
        title = "{The history of star formation in a {\ensuremath{\Lambda}} cold dark matter universe}",
      journal = {\mnras},
         year = 2003,
        month = feb,
       volume = {339},
       number = {2},
        pages = {312-334},
          doi = {10.1046/j.1365-8711.2003.06207.x},
archivePrefix = {arXiv},
       eprint = {astro-ph/0206395},
 primaryClass = {astro-ph},
       adsurl = {https://ui.adsabs.harvard.edu/abs/2003MNRAS.339..312S}
}

@ARTICLE{Springel_2005,
       author = {{Springel}, Volker},
        title = "{The cosmological simulation code GADGET-2}",
      journal = {\mnras},
         year = 2005,
        month = dec,
       volume = {364},
       number = {4},
        pages = {1105-1134},
          doi = {10.1111/j.1365-2966.2005.09655.x},
archivePrefix = {arXiv},
       eprint = {astro-ph/0505010},
 primaryClass = {astro-ph},
       adsurl = {https://ui.adsabs.harvard.edu/abs/2005MNRAS.364.1105S}
}

@ARTICLE{Mosconi_2001,
       author = {{Mosconi}, M.~B. and {Tissera}, P.~B. and {Lambas}, D.~G. and {Cora}, S.~A.},
        title = "{Chemical evolution using smooth particle hydrodynamical cosmological simulations - I. Implementation, tests and first results}",
      journal = {\mnras},
         year = 2001,
        month = jul,
       volume = {325},
       number = {1},
        pages = {34-48},
          doi = {10.1046/j.1365-8711.2001.04198.x},
archivePrefix = {arXiv},
       eprint = {astro-ph/0007074},
 primaryClass = {astro-ph},
       adsurl = {https://ui.adsabs.harvard.edu/abs/2001MNRAS.325...34M}
}

@ARTICLE{Scannapieco_2005,
       author = {{Scannapieco}, C. and {Tissera}, P.B. and {White}, S.D.M. and {Springel}, V.},
        title = "{Feedback and metal enrichment in cosmological smoothed particle hydrodynamics simulations - I. A model for chemical enrichment}",
      journal = {\mnras},
         year = 2005,
        month = dec,
       volume = {364},
       number = {2},
        pages = {552-564},
          doi = {10.1111/j.1365-2966.2005.09574.x},
archivePrefix = {arXiv},
       eprint = {astro-ph/0505440},
 primaryClass = {astro-ph},
       adsurl = {https://ui.adsabs.harvard.edu/abs/2005MNRAS.364..552S}
}

@ARTICLE{Scannapieco_2006,
       author = {{Scannapieco}, C. and {Tissera}, P.B. and {White}, S.D.M. and {Springel}, V.},
        title = "{Feedback and metal enrichment in cosmological SPH simulations - II. A multiphase model with supernova energy feedback}",
      journal = {\mnras},
         year = 2006,
        month = sep,
       volume = {371},
       number = {3},
        pages = {1125-1139},
          doi = {10.1111/j.1365-2966.2006.10785.x},
archivePrefix = {arXiv},
       eprint = {astro-ph/0604524},
 primaryClass = {astro-ph},
       adsurl = {https://ui.adsabs.harvard.edu/abs/2006MNRAS.371.1125S}
}

@ARTICLE{Davis_1985,
       author = {{Davis}, M. and {Efstathiou}, G. and {Frenk}, C.~S. and {White}, S.~D.~M.},
        title = "{The evolution of large-scale structure in a universe dominated by cold dark matter}",
      journal = {\apj},
         year = 1985,
        month = may,
       volume = {292},
        pages = {371-394},
          doi = {10.1086/163168},
       adsurl = {https://ui.adsabs.harvard.edu/abs/1985ApJ...292..371D}
}

@ARTICLE{Dolag_2009,
       author = {{Dolag}, K. and {Borgani}, S. and {Murante}, G. and {Springel}, V.},
        title = "{Substructures in hydrodynamical cluster simulations}",
      journal = {\mnras},
         year = 2009,
        month = oct,
       volume = {399},
       number = {2},
        pages = {497-514},
          doi = {10.1111/j.1365-2966.2009.15034.x},
archivePrefix = {arXiv},
       eprint = {0808.3401},
 primaryClass = {astro-ph},
       adsurl = {https://ui.adsabs.harvard.edu/abs/2009MNRAS.399..497D}
}

@ARTICLE{Knollmann_2009,
       author = {{Knollmann}, Steffen R. and {Knebe}, Alexander},
        title = "{AHF: Amiga's Halo Finder}",
      journal = {\apjs},
         year = 2009,
        month = jun,
       volume = {182},
       number = {2},
        pages = {608-624},
          doi = {10.1088/0067-0049/182/2/608},
archivePrefix = {arXiv},
       eprint = {0904.3662},
 primaryClass = {astro-ph.CO},
       adsurl = {https://ui.adsabs.harvard.edu/abs/2009ApJS..182..608K}
}

@ARTICLE{Munoz-escobar_2025,
       author = {{Mu{\~n}oz-Escobar}, Ignacio and {Tissera}, Patricia B. and {Gonzalez-Jara}, Jenny and {Sillero}, Emanuel and {Miranda}, Valentina P. and {Pedrosa}, Susana and {Bignone}, Lucas},
        title = "{The mass-metallicity relation of bulges}",
      journal = {\aap},
         year = 2026,
        month = jan,
       volume = {705},
          eid = {A87},
        pages = {A87},
          doi = {10.1051/0004-6361/202557133},
archivePrefix = {arXiv},
       eprint = {2510.25642},
 primaryClass = {astro-ph.GA},
       adsurl = {https://ui.adsabs.harvard.edu/abs/2026A&A...705A..87M}
}

@ARTICLE{Tapia_2022,
       author = {{Tapia}, B. and {Tissera}, P.~B. and {Sillero}, E. and {Casanueva}, C. and {Pedrosa}, S. and {Bignone}, L. and {Dominguez Tenreiro}, R. and {Padilla}, N.},
        title = "{Insight into the physical processes that shape the metallicity profiles in galaxies}",
      journal = {Boletin de la Asociacion Argentina de Astronomia La Plata Argentina},
         year = 2022,
        month = jul,
       volume = {63},
        pages = {256-258},
       adsurl = {https://ui.adsabs.harvard.edu/abs/2022BAAA...63..256T}
}

@ARTICLE{Tapia-contreras_2025,
       author = {{Tapia-Contreras}, Brian and {Tissera}, Patricia B. and {Sillero}, Emanuel and {Gonzalez-Jara}, Jenny and {Casanueva-Villarreal}, Catalina and {Pedrosa}, Susana and {Bignone}, Lucas and {Padilla}, Nelson D. and {Dom{\'\i}nguez-Tenreiro}, Rosa},
        title = "{Insight into the physical processes that shape the metallicity profiles in galaxies}",
      journal = {\aap},
         year = 2025,
        month = aug,
       volume = {700},
          eid = {A69},
        pages = {A69},
          doi = {10.1051/0004-6361/202554013},
archivePrefix = {arXiv},
       eprint = {2502.02080},
 primaryClass = {astro-ph.GA},
       adsurl = {https://ui.adsabs.harvard.edu/abs/2025A&A...700A..69T}
}

@ARTICLE{Rodriguez_2022,
       author = {{Rodr{\'\i}guez}, S. and {Garcia Lambas}, D. and {Padilla}, N.~D. and {Tissera}, P. and {Bignone}, L. and {Dominguez-Tenreiro}, R. and {Gonzalez}, R. and {Pedrosa}, S.},
        title = "{Satellite galaxies in groups in the CIELO Project I. Gas removal from galaxies and its re-distribution in the intragroup medium}",
      journal = {\mnras},
         year = 2022,
        month = aug,
       volume = {514},
       number = {4},
        pages = {6157-6172},
          doi = {10.1093/mnras/stac1377},
archivePrefix = {arXiv},
       eprint = {2205.06886},
 primaryClass = {astro-ph.GA},
       adsurl = {https://ui.adsabs.harvard.edu/abs/2022MNRAS.514.6157R}
}

@ARTICLE{Cataldi_2023,
       author = {{Cataldi}, P. and {Pedrosa}, S.~E. and {Tissera}, P.~B. and {Artale}, M.~C. and {Padilla}, N.~D. and {Dominguez-Tenreiro}, R. and {Bignone}, L. and {Gonzalez}, R. and {Pellizza}, L.~J.},
        title = "{Redshift evolution of the dark matter haloes shapes}",
      journal = {\mnras},
         year = 2023,
        month = aug,
       volume = {523},
       number = {2},
        pages = {1919-1932},
          doi = {10.1093/mnras/stad1601},
archivePrefix = {arXiv},
       eprint = {2302.08853},
 primaryClass = {astro-ph.GA},
       adsurl = {https://ui.adsabs.harvard.edu/abs/2023MNRAS.523.1919C}
}

@ARTICLE{Casanueva_2024,
       author = {{Casanueva-Villarreal}, C. and {Tissera}, P.~B. and {Padilla}, N. and {Liu}, B. and {Bromm}, V. and {Pedrosa}, S. and {Bignone}, L. and {Dominguez-Tenreiro}, R.},
        title = "{Impact of primordial black hole dark matter on gas properties at very high redshift: Semianalytical model}",
      journal = {\aap},
         year = 2024,
        month = aug,
       volume = {688},
          eid = {A183},
        pages = {A183},
          doi = {10.1051/0004-6361/202449650},
archivePrefix = {arXiv},
       eprint = {2405.02206},
 primaryClass = {astro-ph.GA},
       adsurl = {https://ui.adsabs.harvard.edu/abs/2024A&A...688A.183C}
}

@ARTICLE{Tissera_2025,
       author = {{Tissera}, Patricia B. and {Bignone}, Lucas and {Gonzalez-Jara}, Jenny and {Mu{\~n}oz-Escobar}, Ignacio and {Cataldi}, Pedro and {Miranda}, Valentina P. and {Barrientos-Acevedo}, Daniela and {Tapia-Contreras}, Brian and {Pedrosa}, Susana and {Padilla}, Nelson and {Dominguez-Tenreiro}, Rosa and {Casanueva-Villarreal}, Catalina and {Sillero}, Emanuel and {Silva-Mella}, Benjamin and {Shailesh}, Isha and {Jara-Ferreira}, Francisco},
        title = "{The CIELO project: The chemo-dynamical properties of galaxies and the cosmic web}",
      journal = {\aap},
         year = 2025,
        month = may,
       volume = {697},
          eid = {A134},
        pages = {A134},
          doi = {10.1051/0004-6361/202453348},
archivePrefix = {arXiv},
       eprint = {2501.05978},
 primaryClass = {astro-ph.GA},
       adsurl = {https://ui.adsabs.harvard.edu/abs/2025A&A...697A.134T}
}

@ARTICLE{Nomoto_2013,
       author = {{Nomoto}, Ken'ichi and {Kobayashi}, Chiaki and {Tominaga}, Nozomu},
        title = "{Nucleosynthesis in Stars and the Chemical Enrichment of Galaxies}",
      journal = {\araa},
         year = 2013,
        month = aug,
       volume = {51},
       number = {1},
        pages = {457-509},
          doi = {10.1146/annurev-astro-082812-140956},
       adsurl = {https://ui.adsabs.harvard.edu/abs/2013ARA&A..51..457N}
}

@ARTICLE{Maiolino_2019,
       author = {{Maiolino}, R. and {Mannucci}, F.},
        title = "{De re metallica: the cosmic chemical evolution of galaxies}",
      journal = {\aapr},
         year = 2019,
        month = feb,
       volume = {27},
       number = {1},
          eid = {3},
        pages = {3},
          doi = {10.1007/s00159-018-0112-2},
archivePrefix = {arXiv},
       eprint = {1811.09642},
 primaryClass = {astro-ph.GA},
       adsurl = {https://ui.adsabs.harvard.edu/abs/2019A&ARv..27....3M}
}

@ARTICLE{Gallazzi_2005,
       author = {{Gallazzi}, Anna and {Charlot}, St{\'e}phane and {Brinchmann}, Jarle and {White}, Simon D.~M. and {Tremonti}, Christy A.},
        title = "{The ages and metallicities of galaxies in the local universe}",
      journal = {\mnras},
         year = 2005,
        month = sep,
       volume = {362},
       number = {1},
        pages = {41-58},
          doi = {10.1111/j.1365-2966.2005.09321.x},
archivePrefix = {arXiv},
       eprint = {astro-ph/0506539},
 primaryClass = {astro-ph},
       adsurl = {https://ui.adsabs.harvard.edu/abs/2005MNRAS.362...41G}
}

@ARTICLE{Licquia_2015,
       author = {{Licquia}, Timothy C. and {Newman}, Jeffrey A.},
        title = "{Improved Estimates of the Milky Way's Stellar Mass and Star Formation Rate from Hierarchical Bayesian Meta-Analysis}",
      journal = {\apj},
         year = 2015,
        month = jun,
       volume = {806},
       number = {1},
          eid = {96},
        pages = {96},
          doi = {10.1088/0004-637X/806/1/96},
archivePrefix = {arXiv},
       eprint = {1407.1078},
 primaryClass = {astro-ph.GA},
       adsurl = {https://ui.adsabs.harvard.edu/abs/2015ApJ...806...96L}
}

@INPROCEEDINGS{Sick_2015,
       author = {{Sick}, Jonathan and {Courteau}, Stephane and {Cuillandre}, Jean-Charles and {Dalcanton}, Julianne and {de Jong}, Roelof and {McDonald}, Michael and {Simard}, Dana and {Tully}, R. Brent},
        title = "{The Stellar Mass of M31 as inferred by the Andromeda Optical \& Infrared Disk Survey}",
    booktitle = {Galaxy Masses as Constraints of Formation Models},
         year = 2015,
       editor = {{Cappellari}, Michele and {Courteau}, St{\'e}phane},
       series = {IAU Symposium},
       volume = {311},
        month = apr,
        pages = {82-85},
          doi = {10.1017/S1743921315003440},
archivePrefix = {arXiv},
       eprint = {1410.0017},
 primaryClass = {astro-ph.GA},
       adsurl = {https://ui.adsabs.harvard.edu/abs/2015IAUS..311...82S}}

@ARTICLE{Fall_2018,
       author = {{Romanowsky}, Aaron J. and {Fall}, S. Michael},
        title = "{Angular Momentum and Galaxy Formation Revisited}",
      journal = {\apjs},
         year = 2012,
        month = dec,
       volume = {203},
       number = {2},
          eid = {17},
        pages = {17},
          doi = {10.1088/0067-0049/203/2/17},
archivePrefix = {arXiv},
       eprint = {1207.4189},
 primaryClass = {astro-ph.CO},
       adsurl = {https://ui.adsabs.harvard.edu/abs/2012ApJS..203...17R}
}

@ARTICLE{Tsakonas_2025,
       author = {{Tsakonas}, C. and {Arnaboldi}, M. and {Bhattacharya}, S. and {Hammer}, F. and {Yang}, Y. and {Gerhard}, O. and {Wyse}, R.~F.~G. and {Hatzidimitriou}, D.},
        title = "{The survey of planetary nebulae in Andromeda (M31): VII. Predictions of a major merger simulation model compared with chemodynamical data of the disc and inner halo substructures}",
      journal = {\aap},
         year = 2025,
        month = jul,
       volume = {699},
          eid = {A56},
        pages = {A56},
          doi = {10.1051/0004-6361/202453175},
archivePrefix = {arXiv},
       eprint = {2502.00886},
 primaryClass = {astro-ph.GA},
       adsurl = {https://ui.adsabs.harvard.edu/abs/2025A&A...699A..56T}
}

@ARTICLE{Dsouza_M31_2018,
       author = {{D'Souza}, Richard and {Bell}, Eric F.},
        title = "{The Andromeda galaxy's most important merger about 2 billion years ago as M32's likely progenitor}",
      journal = {Nature Astronomy},
         year = 2018,
        month = jul,
       volume = {2},
        pages = {737-743},
          doi = {10.1038/s41550-018-0533-x},
archivePrefix = {arXiv},
       eprint = {1807.08819},
 primaryClass = {astro-ph.GA},
       adsurl = {https://ui.adsabs.harvard.edu/abs/2018NatAs...2..737D}
}

@ARTICLE{Escala_2020,
       author = {{Escala}, Ivanna and {Gilbert}, Karoline M. and {Kirby}, Evan N. and {Wojno}, Jennifer and {Cunningham}, Emily C. and {Guhathakurta}, Puragra},
        title = "{Elemental Abundances in M31: A Comparative Analysis of Alpha and Iron Element Abundances in the the Outer Disk, Giant Stellar Stream, and Inner Halo of M31}",
      journal = {\apj},
         year = 2020,
        month = feb,
       volume = {889},
       number = {2},
          eid = {177},
        pages = {177},
          doi = {10.3847/1538-4357/ab6659},
archivePrefix = {arXiv},
       eprint = {1909.00006},
 primaryClass = {astro-ph.GA},
       adsurl = {https://ui.adsabs.harvard.edu/abs/2020ApJ...889..177E}
}

@ARTICLE{Hammer_2018,
       author = {{Hammer}, F. and {Yang}, Y.~B. and {Wang}, J.~L. and {Ibata}, R. and {Flores}, H. and {Puech}, M.},
        title = "{A 2-3 billion year old major merger paradigm for the Andromeda galaxy and its outskirts}",
      journal = {\mnras},
         year = 2018,
        month = feb,
       volume = {475},
       number = {2},
        pages = {2754-2767},
          doi = {10.1093/mnras/stx3343},
archivePrefix = {arXiv},
       eprint = {1801.04279},
 primaryClass = {astro-ph.GA},
       adsurl = {https://ui.adsabs.harvard.edu/abs/2018MNRAS.475.2754H}
}

@ARTICLE{Durrell_2010,
       author = {{Durrell}, Patrick R. and {Sarajedini}, Ata and {Chandar}, Rupali},
        title = "{Deep HST/ACS Photometry of the M81 Halo}",
      journal = {\apj},
         year = 2010,
        month = aug,
       volume = {718},
       number = {2},
        pages = {1118-1127},
          doi = {10.1088/0004-637X/718/2/1118},
archivePrefix = {arXiv},
       eprint = {1006.2036},
 primaryClass = {astro-ph.CO},
       adsurl = {https://ui.adsabs.harvard.edu/abs/2010ApJ...718.1118D}
}

@ARTICLE{Fragkoudi_2025,
       author = {{Fragkoudi}, Francesca and {Grand}, Robert J.~J. and {Pakmor}, R{\"u}diger and {G{\'o}mez}, Facundo and {Marinacci}, Federico and {Springel}, Volker},
        title = "{Bar formation and evolution in the cosmological context: inputs from the Auriga simulations}",
      journal = {\mnras},
         year = 2025,
        month = apr,
       volume = {538},
       number = {3},
        pages = {1587-1608},
          doi = {10.1093/mnras/staf389},
archivePrefix = {arXiv},
       eprint = {2406.09453},
 primaryClass = {astro-ph.GA},
       adsurl = {https://ui.adsabs.harvard.edu/abs/2025MNRAS.538.1587F}
}

\begin{appendix}
\onecolumn
\section{Cardiograms}\label{appendix:all_cardiograms}

We present cardiograms for the remaining 23 stellar halos studied in this article in Fig.~\ref{fig:cardiogram_all}. We keep the same labels as Fig.~\ref{fig:example_cardiogram}. Galaxies are ordered by stellar galaxy mass, with the most massive at the top and the least massive at the bottom. Some cardiograms do not extend to 12~Gyr due to our numerical resolution criterion; values are omitted when this criterion is not met.

\begin{figure}[h!]
    \centering
    \includegraphics[width=0.63\linewidth]{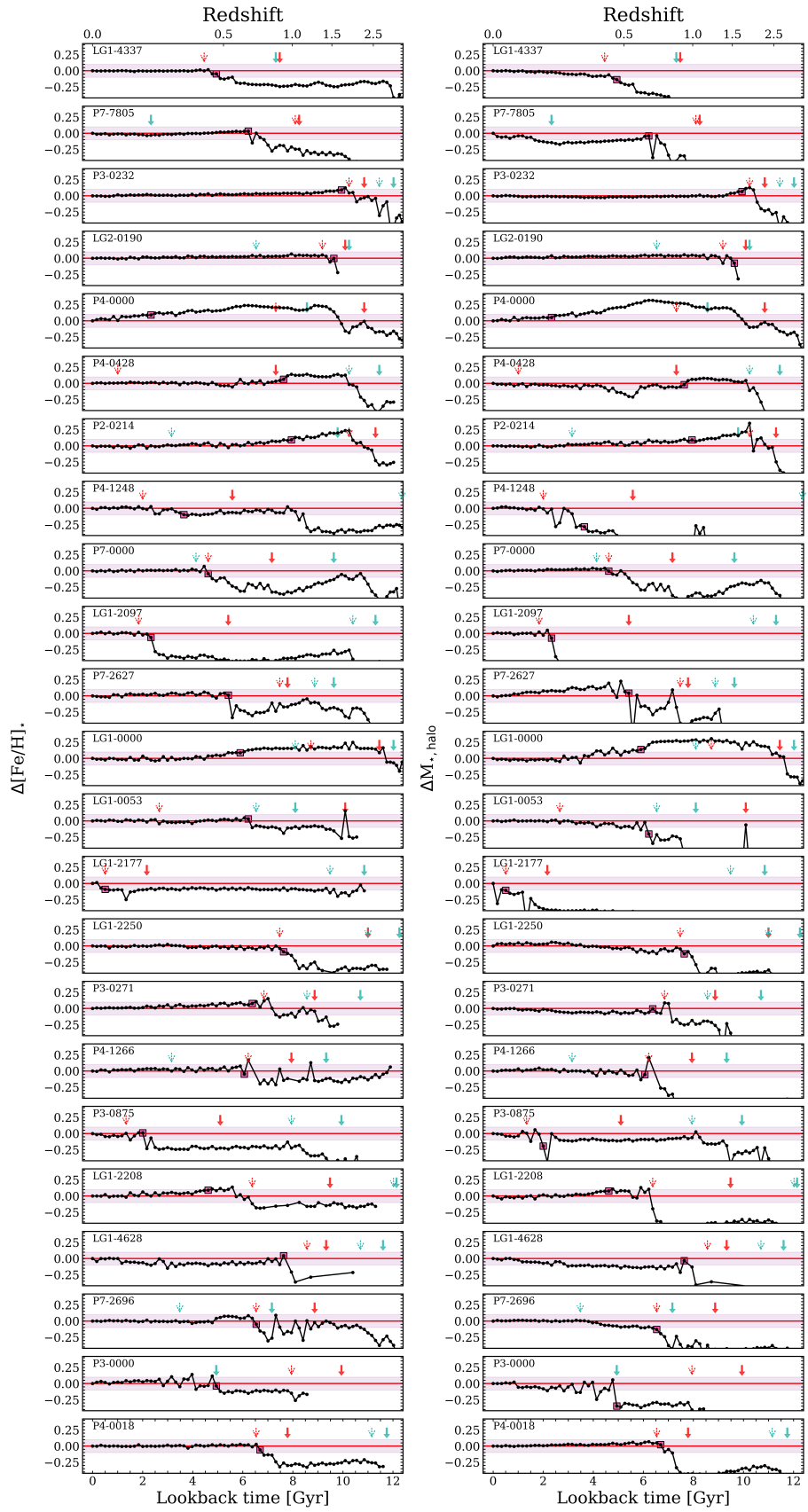}
    \caption{Cardiograms for the remaining 23 stellar halos analyzed in this article. Same labels as Fig.~\ref{fig:example_cardiogram}.}
    \label{fig:cardiogram_all}
\end{figure}

\section{Stellar halo mass - $\rm t_{90}$}\label{appendix:SH-t90}

In Sec.~\ref{sec:observations}, we discuss observational measurements that can be used to estimate $\tst$. We report a positive correlation between $\tst$ and $\tninety$, i.e the time before which 90 percent of the halo stars were born. Estimating $\tninety$ of the stellar halo from observations is challenging; therefore, we also explore its relation with stellar mass. Figure~\ref{fig:appedix_Mhalo_t90} shows $\tninety$ as a function of stellar halo mass for three definitions: ex-situ halo stars (left panel), accreted halo stars (middle panel), and all halo populations (right panel). For comparison, we include the results of \citet{Harmsen_2023}, who report an anti-correlation between $\tninety$ and accreted stellar halo mass in the TNG50 simulation. Their relation exhibits substantial scatter at the low-mass end. A direct comparison with our results is not straightforward, as the two studies probe different mass ranges and adopt different definitions of the accreted stellar population. \citet{Harmsen_2023} analyze galaxies with stellar masses in the range $3\times 10^{10} - 15 \times 10^{10} \rm M_\odot$, whereas CIELO sample spans from $1.6\times 10^{9} - 5.5 \times 10^{10} \rm M_\odot$, only three CIELO galaxies fall in the mass range selected by \citet{Harmsen_2023}. 

Fig.~\ref{fig:appedix_Mhalo_t90} shows that when only ex-situ stars are considered, the stellar halo populations tend to be older and the relation becomes more consistent with the anti-correlation reported by \citet{Harmsen_2023}. However, including endo-debris stars modifies the trend and although it is within the scatter shown by TNG50 galaxies, this reflects the role of the subgrid physics implemented in the simulations. Particularly the regulation of gas in satellites and their subsequent star formation histories. In our simulations, gas-rich satellites can continue forming stars after infall (endo-debris stars), producing younger stellar populations that contribute to the stellar halo. At lower stellar masses, where the TNG50 relation already exhibits substantial scatter, the CIELO halos fall within this dispersion. These results suggest that stellar feedback and the regulation of star formation in satellite galaxies may provide a useful test of subgrid physics once larger statistical samples of stellar halo assembly times are available.

\begin{figure}[h!]
    \centering
    \includegraphics[width=0.9\linewidth]{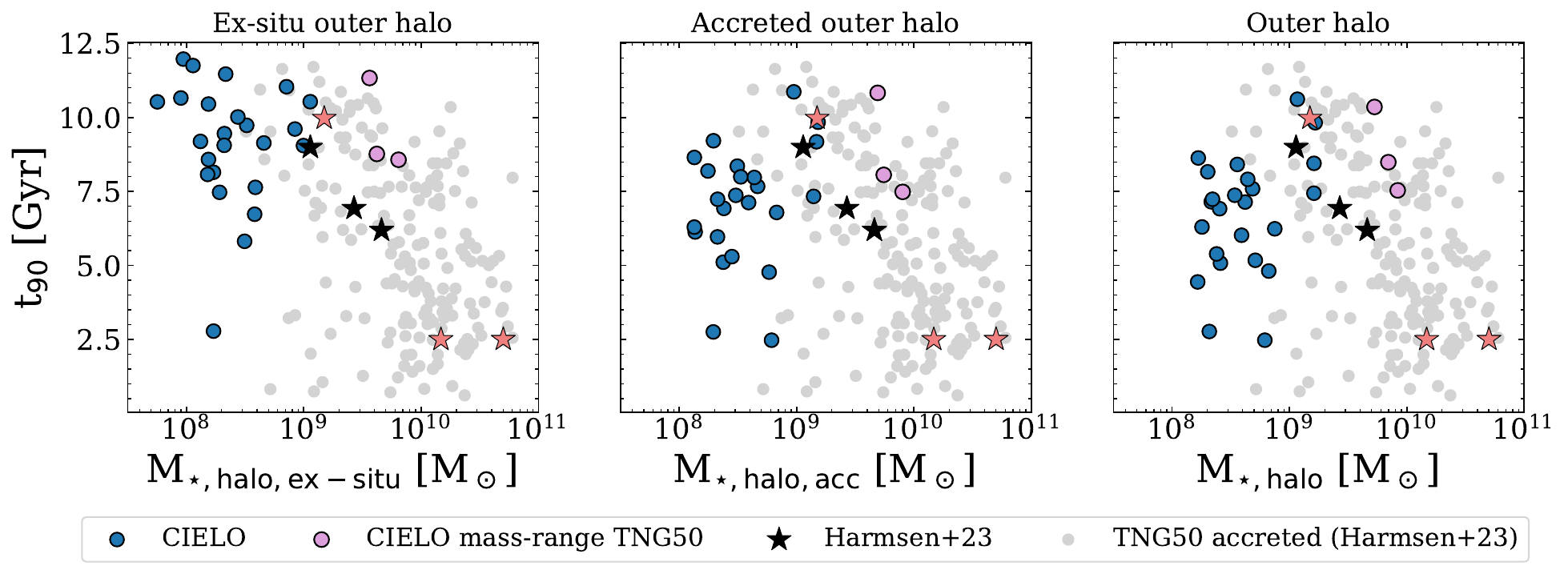}
    \caption{$\tninety$ as function of the stellar halo mass, quantities in each panel are estimated considering ex-situ stars (left), accreted stars (middle) and all the populations, i.e. ex-situ, endo-debris and in-situ (right). Pink circles are the three CIELO galaxies that overlap with TNG50 sample selection. Stars are values reported from \citet{Harmsen_2023} using the AGB/RGB ratio (black) and constrains on the star formation history of stellar halos reported in the literature (orange), further details can be found in section 5.4 of \citet{Harmsen_2023}. Gray circles are TNG50 sample from \citet{Harmsen_2023}.}
    \label{fig:appedix_Mhalo_t90}
\end{figure}

\end{appendix}
\end{document}